\documentclass[JHEP,12pt]{article}

\usepackage{graphics,color,soul}
\usepackage{graphicx}
\usepackage{amssymb}
\usepackage{jheppub}
\usepackage{lmodern,mathtools}
\usepackage{caption}
\usepackage{subcaption}
\usepackage{placeins}
\usepackage{booktabs}
\usepackage[english]{babel}
\usepackage{amsmath,amssymb,amsbsy,amstext, amsthm, simplewick}
\usepackage{hyperref}
\usepackage{tikz}
\usetikzlibrary{decorations.markings}
\usepackage{xcolor}
\usepackage[export]{adjustbox}
\usepackage{caption}
\usepackage{amsfonts}
\usepackage{amssymb}
\usepackage{upgreek}
\usepackage{simplewick}
 \usepackage{exscale,relsize}
\usepackage{mathtools}
\usepackage{comment}
\usepackage{cleveref}

\usepackage{graphicx}
\usepackage[margin=0cm,labelfont={sf,bf,footnotesize},textfont={sf,footnotesize}]{caption}

\def\be#1\ee{\begin{align}#1\end{align}}

\begin{document}

\title{Quantum gravity on the bulk cone}
\author[a,c] {Ahmed Almheiri,}
\author[b]{ Matthew Dodelson,}
\author[c]{and Zohar Komargodski}
\affiliation[a]{New York University Abu Dhabi, \\ Abu Dhabi, P.O. Box 129188, United Arab Emirates }
\affiliation[b]{Center for the Fundamental Laws of Nature, Harvard University,\\
17 Oxford Street, Cambridge, MA 02138, USA}
\affiliation[c]{Simons Center for Geometry and Physics, Stony Brook University,\\
Stony Brook, NY 11794, USA}
\emailAdd{almheiri@nyu.edu}
\emailAdd{mdodelson@fas.harvard.edu}
\emailAdd{zkomargo@gmail.com}

\abstract{Bulk-cone singularities in holographic correlators encode propagation along bulk null geodesics. The prototypical example is an AdS black hole, where incoming null geodesics can wind around the photon sphere and return to the boundary.
These singularities appear in real time in the thermal two-point function at {\it finite volume}. We show that conformal field theories dual to gravitational theories at finite $G_N$ cannot have such singularities.  We rule out these singularities both for individual microstates and for the thermal state. 
The main thrust of the argument is to invoke an analytic continuation in the temperature. As a microscopic example, we construct an AdS$_3$ geometry with two colliding shocks which exhibits a bulk cone singularity. We show by an explicit calculation in the CFT that this singularity is smoothed out at finite $G_N$. Qualitatively, the singularity resolution is a consequence of the fuzziness of the bulk solution in quantum gravity.}

\hypersetup{
  pdftitle={Quantum gravity on the bulk cone},
  pdfauthor={Ahmed Almheiri, Matthew Dodelson, and Zohar Komargodski}
}
\maketitle

\section{Introduction}
\label{sec:introduction}

 A light ray in AdS can depart from the boundary and come back in finite time. Its return leaves  a singularity in the two-point function at the time of arrival. In the presence of a black hole, a null geodesic can spend a long time near the photon sphere (never entering the horizon) before escaping, and the resulting singularity generally occurs at a time-like separation on the boundary.
This may already sound surprising: since the boundary theory is a local conformal field theory, a time-like singularity might appear out of the ordinary. 

 Such singularities at time-like separation were considered by Hubeny, Liu and Rangamani, who coined the term \emph{bulk-cone singularity}~\cite{HubenyLiuRangamani:2006}. Tidal effects on the string worldsheet near the bulk-cone singularity were investigated in~\cite{DodelsonOoguri:2020,Sheta:StringyPhotonSphere}. More recently, ~\cite{DodelsonEtAl:2023} computed the leading form of the black-hole bulk-cone singularity, including its strength and its relation to the instability of the photon sphere. The effect of backreaction of the geodesic on the singularity was studied in \cite{Batra:2026otu}.
 
 We will briefly review the classical gravity prediction, and then turn to the constraints imposed by a CFT at finite $G_N$.

 Consider a four-dimensional CFT on $S^3_R\times\mathbb R$, with a Hermitian scalar primary $\phi$ of dimension $\Delta_\phi$, normalized by $\langle\phi(x)\phi(0)\rangle_{\mathbb R^4}=|x|^{-2\Delta_\phi}$. We use the thermal Wightman function
 \begin{align} G_\beta(t,\theta)
={1\over Z}\operatorname{Tr}\!\left(e^{-\beta H}\phi(t,n)\phi(0,n_0)\right),
\label{eq:intro-wightman}
 \end{align}
 where $n, n_0$ are unit vectors on the $S^3$. 
We take $t>0$ to have proper time ordering. 
 
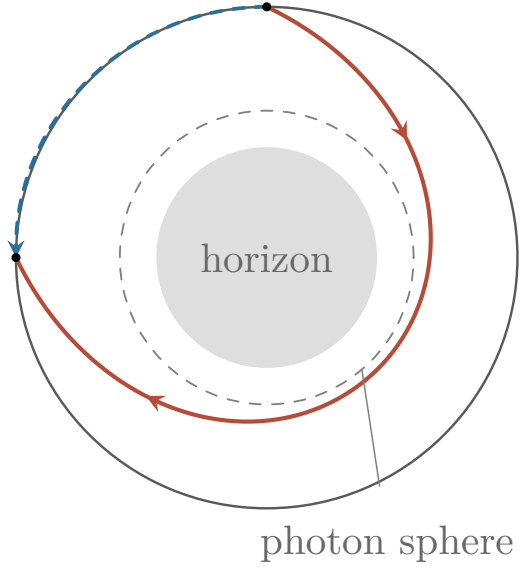
\begin{figure}[!htbp]
 \centering
 \definecolor{bulkRay}{RGB}{182,75,56}
 \definecolor{boundaryRay}{RGB}{43,111,155}
 \resizebox{0.44\textwidth}{!}{%
 \begin{tikzpicture}[font=\small,>=stealth,line cap=round,line join=round]
 \begin{scope}[shift={(3,3.05)},scale=2.35]
  \draw[black!65,line width=.65pt] (0,0) circle (1);
 \fill[black!13] (0,0) circle (.440137);
 \draw[black!50,dashed,line width=.5pt] (0,0) circle (.585786);
 \draw[bulkRay,line width=1.1pt] plot coordinates {(0.00000,1.00000) (0.01551,0.99247) (0.03079,0.98481) (0.04584,0.97702) (0.06066,0.96911) (0.07526,0.96109) (0.08964,0.95295) (0.10381,0.94470) (0.11776,0.93634) (0.13150,0.92788) (0.14503,0.91931) (0.15835,0.91065) (0.17147,0.90189) (0.18439,0.89303) (0.19710,0.88409) (0.20962,0.87505) (0.22195,0.86592) (0.23408,0.85671) (0.24602,0.84741) (0.25777,0.83803) (0.26934,0.82857) (0.28072,0.81903) (0.29191,0.80941) (0.30293,0.79971) (0.31377,0.78993) (0.32443,0.78008) (0.33491,0.77016) (0.34522,0.76016) (0.35536,0.75009) (0.36533,0.73995) (0.37512,0.72974) (0.38475,0.71945) (0.39421,0.70910) (0.40350,0.69868) (0.41263,0.68819) (0.42159,0.67763) (0.43039,0.66700) (0.43903,0.65631) (0.44750,0.64554) (0.45581,0.63471) (0.46396,0.62382) (0.47195,0.61285) (0.47978,0.60182) (0.48745,0.59072) (0.49495,0.57956) (0.50230,0.56833) (0.50949,0.55703) (0.51652,0.54567) (0.52338,0.53424) (0.53009,0.52274) (0.53663,0.51117) (0.54302,0.49954) (0.54924,0.48785) (0.55529,0.47608) (0.56119,0.46426) (0.56692,0.45236) (0.57248,0.44040) (0.57788,0.42837) (0.58311,0.41628) (0.58816,0.40412) (0.59305,0.39189) (0.59777,0.37960) (0.60231,0.36725) (0.60668,0.35483) (0.61087,0.34234) (0.61488,0.32980) (0.61872,0.31719) (0.62237,0.30451) (0.62583,0.29178) (0.62911,0.27899) (0.63220,0.26613) (0.63510,0.25322) (0.63780,0.24025) (0.64031,0.22722) (0.64262,0.21414) (0.64472,0.20101) (0.64663,0.18782) (0.64833,0.17458) (0.64981,0.16130) (0.65109,0.14797) (0.65215,0.13459) (0.65299,0.12118) (0.65362,0.10772) (0.65402,0.09423) (0.65419,0.08070) (0.65414,0.06715) (0.65385,0.05357) (0.65333,0.03996) (0.65257,0.02633) (0.65157,0.01269) (0.65033,-0.00097) (0.64884,-0.01463) (0.64710,-0.02830) (0.64511,-0.04197) (0.64287,-0.05563) (0.64038,-0.06928) (0.63763,-0.08292) (0.63461,-0.09654) (0.63134,-0.11013) (0.62781,-0.12369) (0.62400,-0.13721) (0.61994,-0.15069) (0.61561,-0.16411) (0.61100,-0.17749) (0.60614,-0.19080) (0.60100,-0.20404) (0.59559,-0.21720) (0.58991,-0.23028) (0.58397,-0.24327) (0.57776,-0.25617) (0.57128,-0.26896) (0.56453,-0.28164) (0.55752,-0.29420) (0.55025,-0.30664) (0.54271,-0.31894) (0.53492,-0.33111) (0.52687,-0.34313) (0.51857,-0.35499) (0.51002,-0.36669) (0.50123,-0.37823) (0.49219,-0.38959) (0.48292,-0.40077) (0.47341,-0.41176) (0.46367,-0.42255) (0.45371,-0.43315) (0.44354,-0.44354) (0.43315,-0.45371) (0.42255,-0.46367) (0.41176,-0.47341) (0.40077,-0.48292) (0.38959,-0.49219) (0.37823,-0.50123) (0.36669,-0.51002) (0.35499,-0.51857) (0.34313,-0.52687) (0.33111,-0.53492) (0.31894,-0.54271) (0.30664,-0.55025) (0.29420,-0.55752) (0.28164,-0.56453) (0.26896,-0.57128) (0.25617,-0.57776) (0.24327,-0.58397) (0.23028,-0.58991) (0.21720,-0.59559) (0.20404,-0.60100) (0.19080,-0.60614) (0.17749,-0.61100) (0.16411,-0.61561) (0.15069,-0.61994) (0.13721,-0.62400) (0.12369,-0.62781) (0.11013,-0.63134) (0.09654,-0.63461) (0.08292,-0.63763) (0.06928,-0.64038) (0.05563,-0.64287) (0.04197,-0.64511) (0.02830,-0.64710) (0.01463,-0.64884) (0.00097,-0.65033) (-0.01269,-0.65157) (-0.02633,-0.65257) (-0.03996,-0.65333) (-0.05357,-0.65385) (-0.06715,-0.65414) (-0.08070,-0.65419) (-0.09423,-0.65402) (-0.10772,-0.65362) (-0.12118,-0.65299) (-0.13459,-0.65215) (-0.14797,-0.65109) (-0.16130,-0.64981) (-0.17458,-0.64833) (-0.18782,-0.64663) (-0.20101,-0.64472) (-0.21414,-0.64262) (-0.22722,-0.64031) (-0.24025,-0.63780) (-0.25322,-0.63510) (-0.26613,-0.63220) (-0.27899,-0.62911) (-0.29178,-0.62583) (-0.30451,-0.62237) (-0.31719,-0.61872) (-0.32980,-0.61488) (-0.34234,-0.61087) (-0.35483,-0.60668) (-0.36725,-0.60231) (-0.37960,-0.59777) (-0.39189,-0.59305) (-0.40412,-0.58816) (-0.41628,-0.58311) (-0.42837,-0.57788) (-0.44040,-0.57248) (-0.45236,-0.56692) (-0.46426,-0.56119) (-0.47608,-0.55529) (-0.48785,-0.54924) (-0.49954,-0.54302) (-0.51117,-0.53663) (-0.52274,-0.53009) (-0.53424,-0.52338) (-0.54567,-0.51652) (-0.55703,-0.50949) (-0.56833,-0.50230) (-0.57956,-0.49495) (-0.59072,-0.48745) (-0.60182,-0.47978) (-0.61285,-0.47195) (-0.62382,-0.46396) (-0.63471,-0.45581) (-0.64554,-0.44750) (-0.65631,-0.43903) (-0.66700,-0.43039) (-0.67763,-0.42159) (-0.68819,-0.41263) (-0.69868,-0.40350) (-0.70910,-0.39421) (-0.71945,-0.38475) (-0.72974,-0.37512) (-0.73995,-0.36533) (-0.75009,-0.35536) (-0.76016,-0.34522) (-0.77016,-0.33491) (-0.78008,-0.32443) (-0.78993,-0.31377) (-0.79971,-0.30293) (-0.80941,-0.29191) (-0.81903,-0.28072) (-0.82857,-0.26934) (-0.83803,-0.25777) (-0.84741,-0.24602) (-0.85671,-0.23408) (-0.86592,-0.22195) (-0.87505,-0.20962) (-0.88409,-0.19710) (-0.89303,-0.18439) (-0.90189,-0.17147) (-0.91065,-0.15835) (-0.91931,-0.14503) (-0.92788,-0.13150) (-0.93634,-0.11776) (-0.94470,-0.10381) (-0.95295,-0.08964) (-0.96109,-0.07526) (-0.96911,-0.06066) (-0.97702,-0.04584) (-0.98481,-0.03079) (-0.99247,-0.01551) (-1.00000,0.00000)};
 \draw[boundaryRay,dashed,line width=1pt,->] (0,1) arc (90:180:1);
 \draw[bulkRay,line width=1.1pt,->] (0.53663,0.51117) -- (0.55973,0.46722);
 \draw[bulkRay,line width=1.1pt,->] (-0.43740,-0.57384) -- (-0.48197,-0.55229);
 \fill (0,1) circle (.018);
 \fill (-1,0) circle (.018);
 \node[text=black!60] at (0,0) {horizon};
 \draw[black!50,line width=.4pt] (.38,-.45) -- (.45,-.91);
 \node[text=black!60] at (.48,-1.14) {photon sphere};
\end{scope}
\end{tikzpicture}%
}
\caption{A null ray in Schwarzschild--AdS$_5$, shown in red. The dashed line is the photon sphere, which is the smallest radius that an ingoing null ray can reach before escaping. The shaded region is the black hole interior, which is inside the photon sphere.}
\label{fig:bulk-cone-geometry}
\end{figure}

 The bulk geometry of interest is Schwarzschild--AdS$_5$, \begin{align}
ds^2&=L^2\left[-f(y)d t^{\,2}+\frac{dy^2}{f(y)}+y^2d\Omega_3^2\right],
  & f(y)&=1+y^2-\frac{y_h^2(1+y_h^2)}{y^2}~.
\label{eq:intro-metric}
 \end{align}
Here $y_h$ is the horizon radius. The temperature of this black hole is
\begin{align}
\beta^{-1}=\frac{1+2y_h^2}{2\pi y_h}~.
\label{eq:intro-temperature}
\end{align}
Figure~\ref{fig:bulk-cone-geometry} shows a null ray returning to the boundary after an  excursion through the bulk.

The endpoints of these rays determine the singularity loci. In the coordinates of~\eqref{eq:intro-metric}, let $b$ be the ratio of a ray's conserved angular momentum to its energy, and write $U(y)=f(y)/y^2$. A returning ray turns at the largest root $y_*$ of $b^2U(y_*)=1$, outside the photon sphere; its elapsed time and total angle along a great circle are
\begin{align}
 t_{\rm bulk}(b)&=2\int_{y_*}^{\infty}\frac{dy}{f(y)\sqrt{1-b^2U(y)}},\\
 \Theta(b)&=2\int_{y_*}^{\infty}\frac{b\,dy}{y^2\sqrt{1-b^2U(y)}}.
 \label{eq:intro-return-locus}
\end{align}
$\Theta$ is essentially the angle on a great circle, except that we have to worry about the fundamental domain. We can map any $\Theta$ back to the fundamental domain  $\theta=\cos^{-1} (\cos(\Theta))$.
A light ray through pure AdS$_5$ returns to the antipodal point, with $\Theta=\pi$; the black hole bends it further, and indeed it is easy to prove that~\eqref{eq:intro-return-locus} satisfies $\Theta>\pi$.\footnote{\[
\Theta=2\int_0^1\frac{dx}{\sqrt{1-x^2}\sqrt{1-a(1+x^2)}}
>2\int_0^1\frac{dx}{\sqrt{1-x^2}}=\pi.
\]}

Varying $b$ traces out the possible singularities. Rays that approach the unstable photon orbit can wind many times before escaping~\cite{HubenyLiuRangamani:2006}.

\begin{figure}[!htbp]
\centering
\definecolor{coneCool}{RGB}{182,75,56}
\definecolor{coneMid}{RGB}{43,111,155}
\definecolor{coneHot}{RGB}{112,136,62}
\begin{tikzpicture}[x=9cm,y=2.4cm,font=\small,>=stealth,line cap=round,line join=round]
 \fill[black!4] (0,0) -- (1,0) -- (1,1) -- cycle;
 \foreach \y in {.5,1,1.5,2,2.5,3}{\draw[black!8] (0,\y) -- (1,\y);}
 \begin{scope}
 \clip (0,0) rectangle (1,3);
 \draw[black!55,dashed,line width=.8pt] (1,1) -- (0,2) -- (1,3);
 \draw[coneCool,line width=1pt] plot coordinates {(1.00000,1.00000) (0.96667,1.03295) (0.93333,1.06524) (0.90000,1.09702) (0.86667,1.12839) (0.83333,1.15943) (0.80000,1.19021) (0.76667,1.22076) (0.73333,1.25114) (0.70000,1.28136) (0.66667,1.31147) (0.63333,1.34146) (0.60000,1.37137) (0.56667,1.40121) (0.53333,1.43098) (0.50000,1.46070) (0.46667,1.49038) (0.43333,1.52001) (0.40000,1.54962) (0.36667,1.57920) (0.33333,1.60875) (0.30000,1.63828) (0.26667,1.66780) (0.23333,1.69730) (0.20000,1.72679) (0.16667,1.75627) (0.13333,1.78574) (0.10000,1.81520) (0.06667,1.84465) (0.03333,1.87410) (0.00000,1.90355)};
 \draw[coneCool,line width=1pt] plot coordinates {(0.00000,1.90355) (0.03333,1.93299) (0.06667,1.96242) (0.10000,1.99185) (0.13333,2.02128) (0.16667,2.05071) (0.20000,2.08013) (0.23333,2.10956) (0.26667,2.13898) (0.30000,2.16840) (0.33333,2.19782) (0.36667,2.22723) (0.40000,2.25665) (0.43333,2.28607) (0.46667,2.31548) (0.50000,2.34490) (0.53333,2.37431) (0.56667,2.40373) (0.60000,2.43314) (0.63333,2.46255) (0.66667,2.49197) (0.70000,2.52138) (0.73333,2.55079) (0.76667,2.58021) (0.80000,2.60962) (0.83333,2.63903) (0.86667,2.66844) (0.90000,2.69786) (0.93333,2.72727) (0.96667,2.75668) (1.00000,2.78609)};
 \draw[coneCool,line width=1pt] plot coordinates {(1.00000,2.78609) (0.96667,2.81551) (0.93333,2.84492) (0.90000,2.87433) (0.86667,2.90374) (0.83333,2.93315) (0.80000,2.96256) (0.76667,2.99198) (0.73333,3.02139) (0.70000,3.05080) (0.66667,3.08021) (0.63333,3.10962) (0.60000,3.13904) (0.56667,3.16845) (0.53333,3.19786) (0.50000,3.22727) (0.46667,3.25668) (0.43333,3.28610) (0.40000,3.31551) (0.36667,3.34492) (0.33333,3.37433) (0.30000,3.40374) (0.26667,3.43315) (0.23333,3.46257) (0.20000,3.49198) (0.16667,3.52139) (0.13333,3.55080) (0.10000,3.58021) (0.06667,3.60962) (0.03333,3.63904) (0.00000,3.66845)};
 \draw[coneMid,line width=1pt] plot coordinates {(1.00000,1.00000) (0.96667,1.03316) (0.93333,1.06602) (0.90000,1.09865) (0.86667,1.13107) (0.83333,1.16334) (0.80000,1.19547) (0.76667,1.22750) (0.73333,1.25943) (0.70000,1.29128) (0.66667,1.32307) (0.63333,1.35481) (0.60000,1.38650) (0.56667,1.41815) (0.53333,1.44977) (0.50000,1.48136) (0.46667,1.51293) (0.43333,1.54447) (0.40000,1.57600) (0.36667,1.60752) (0.33333,1.63902) (0.30000,1.67052) (0.26667,1.70200) (0.23333,1.73347) (0.20000,1.76494) (0.16667,1.79640) (0.13333,1.82786) (0.10000,1.85932) (0.06667,1.89077) (0.03333,1.92221) (0.00000,1.95366)};
 \draw[coneMid,line width=1pt] plot coordinates {(0.00000,1.95366) (0.03333,1.98510) (0.06667,2.01654) (0.10000,2.04797) (0.13333,2.07941) (0.16667,2.11085) (0.20000,2.14228) (0.23333,2.17371) (0.26667,2.20514) (0.30000,2.23658) (0.33333,2.26801) (0.36667,2.29944) (0.40000,2.33087) (0.43333,2.36230) (0.46667,2.39372) (0.50000,2.42515) (0.53333,2.45658) (0.56667,2.48801) (0.60000,2.51944) (0.63333,2.55087) (0.66667,2.58229) (0.70000,2.61372) (0.73333,2.64515) (0.76667,2.67658) (0.80000,2.70800) (0.83333,2.73943) (0.86667,2.77086) (0.90000,2.80229) (0.93333,2.83371) (0.96667,2.86514) (1.00000,2.89657)};
 \draw[coneMid,line width=1pt] plot coordinates {(1.00000,2.89657) (0.96667,2.92800) (0.93333,2.95942) (0.90000,2.99085) (0.86667,3.02228) (0.83333,3.05370) (0.80000,3.08513) (0.76667,3.11656) (0.73333,3.14798) (0.70000,3.17941) (0.66667,3.21084) (0.63333,3.24227) (0.60000,3.27369) (0.56667,3.30512) (0.53333,3.33655) (0.50000,3.36797) (0.46667,3.39940) (0.43333,3.43083) (0.40000,3.46225) (0.36667,3.49368) (0.33333,3.52511) (0.30000,3.55654) (0.26667,3.58796) (0.23333,3.61939) (0.20000,3.65082) (0.16667,3.68224) (0.13333,3.71367) (0.10000,3.74510) (0.06667,3.77652) (0.03333,3.80795) (0.00000,3.83938)};
 \draw[coneHot,line width=1pt] plot coordinates {(1.00000,1.00000) (0.96667,1.03329) (0.93333,1.06649) (0.90000,1.09962) (0.86667,1.13270) (0.83333,1.16573) (0.80000,1.19872) (0.76667,1.23168) (0.73333,1.26461) (0.70000,1.29752) (0.66667,1.33041) (0.63333,1.36328) (0.60000,1.39614) (0.56667,1.42899) (0.53333,1.46183) (0.50000,1.49466) (0.46667,1.52748) (0.43333,1.56029) (0.40000,1.59310) (0.36667,1.62591) (0.33333,1.65871) (0.30000,1.69151) (0.26667,1.72430) (0.23333,1.75709) (0.20000,1.78988) (0.16667,1.82267) (0.13333,1.85546) (0.10000,1.88825) (0.06667,1.92103) (0.03333,1.95381) (0.00000,1.98660)};
 \draw[coneHot,line width=1pt] plot coordinates {(0.00000,1.98660) (0.03333,2.01938) (0.06667,2.05216) (0.10000,2.08494) (0.13333,2.11772) (0.16667,2.15050) (0.20000,2.18328) (0.23333,2.21606) (0.26667,2.24884) (0.30000,2.28162) (0.33333,2.31440) (0.36667,2.34718) (0.40000,2.37996) (0.43333,2.41274) (0.46667,2.44551) (0.50000,2.47829) (0.53333,2.51107) (0.56667,2.54385) (0.60000,2.57663) (0.63333,2.60940) (0.66667,2.64218) (0.70000,2.67496) (0.73333,2.70774) (0.76667,2.74052) (0.80000,2.77329) (0.83333,2.80607) (0.86667,2.83885) (0.90000,2.87163) (0.93333,2.90441) (0.96667,2.93718) (1.00000,2.96996)};
 \draw[coneHot,line width=1pt] plot coordinates {(1.00000,2.96996) (0.96667,3.00274) (0.93333,3.03552) (0.90000,3.06829) (0.86667,3.10107) (0.83333,3.13385) (0.80000,3.16663) (0.76667,3.19941) (0.73333,3.23218) (0.70000,3.26496) (0.66667,3.29774) (0.63333,3.33052) (0.60000,3.36329) (0.56667,3.39607) (0.53333,3.42885) (0.50000,3.46163) (0.46667,3.49441) (0.43333,3.52718) (0.40000,3.55996) (0.36667,3.59274) (0.33333,3.62552) (0.30000,3.65829) (0.26667,3.69107) (0.23333,3.72385) (0.20000,3.75663) (0.16667,3.78941) (0.13333,3.82218) (0.10000,3.85496) (0.06667,3.88774) (0.03333,3.92052) (0.00000,3.95329)};
 \draw[black,line width=1.1pt] (0,0) -- (1,1)
   node[pos=.5,above,sloped,font=\footnotesize] {first boundary signal: $t=\theta$};
 \node[black!55,font=\footnotesize] at (.75,.23) {spacelike};
 \end{scope}
 \draw[->,black!70] (0,0) -- (1.04,0);
 \draw[->,black!70] (0,0) -- (0,3.1);
 \foreach \x/\lab in {0/0,.25/{1/4},.5/{1/2},.75/{3/4},1/1}{\draw[black!70] (\x,0) -- (\x,-.03) node[below] {$\lab$};}
 \foreach \y in {.5,1,1.5,2,2.5,3}{\draw[black!70] (0,\y) -- (-.012,\y) node[left] {$\y$};}
 \node at (.5,-.39) {$\theta/\pi$};
 \node[rotate=90] at (-.105,1.5) {$t/\pi$};
 \draw[black!55,dashed,line width=.8pt] (.045,2.98) -- (.11,2.98);
 \node[anchor=west,font=\footnotesize] at (.13,2.98) {later boundary rays};
 \draw[coneCool,line width=1pt] (.045,2.83) -- (.11,2.83);
 \node[anchor=west,font=\footnotesize] at (.13,2.83) {$\beta^{-1}=0.451$};
 \draw[coneMid,line width=1pt] (.045,2.68) -- (.11,2.68);
 \node[anchor=west,font=\footnotesize] at (.13,2.68) {$\beta^{-1}=0.477$};
 \draw[coneHot,line width=1pt] (.045,2.53) -- (.11,2.53);
 \node[anchor=west,font=\footnotesize] at (.13,2.53) {$\beta^{-1}=0.584$};
\end{tikzpicture}
\caption{Bulk-cone singularities on the boundary, with the source at $(t,\theta)=(0,0)$. The solid black line is the shortest path on the boundary; the dashed lines are later boundary null trajectories. The colored curves follow from~\eqref{eq:intro-return-locus} for $y_h=0.75,1,1.5$. Bulk signals never precede the quickest null boundary trajectory, although they  precede the later null boundary trajectories.}
\label{fig:bulk-cone-temperatures}
\end{figure}
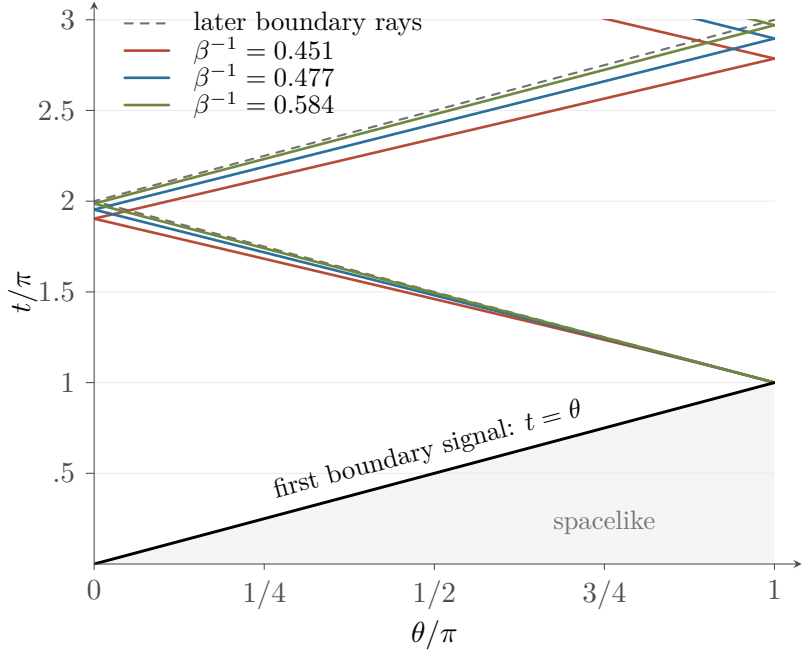

The connection to the boundary Green's function in the thermal state is immediate. The bulk Green's function is singular along null rays, and taking its endpoints to the boundary carries these singularities over to the boundary Wightman function $G_\beta(t,\theta)$. In the classical gravity approximation, we get \begin{equation}\label{bcpower}G_\beta(t,\theta)\sim \frac{1}{\left(t_{\rm bulk}-t+i0\right)^{2\Delta_\phi-3/2}}~,\end{equation}
where we omit a nonsingular prefactor, see~\cite{DodelsonEtAl:2023}. These returns lie in the boundary timelike domain, and generically occur away from the boundary null separations $t=2\pi k\pm\theta$. Figure~\ref{fig:bulk-cone-temperatures} displays these boundary singularities. They have the very important feature that the elapsed angle always satisfies $\Theta>\pi$. The time at which they hit the boundary is shorter than the time it would have taken a light ray to hit the same point had it traversed the same trajectory on the great circle, but it is longer than the shortest possible path for the light ray on the boundary. 
 
Our aim is to constrain such time-like singularities by appealing to the properties of the boundary theory. At finite $G_N$, corresponding to finite $C_T\sim L^3/G_N$, the exact correlator obeys convergence and analyticity conditions that need not commute with the $G_N\to0$ limit. Section~\ref{sec:microstate} applies radial OPE convergence to a Green's function in a microstate, and shows that the singularities must be on the boundary light cone. Therefore, at finite $G_N$, in any given microstate, we cannot have singularities away from the boundary lightcone. 
That is already interesting and strongly suggests that the bulk cone singularities must be resolved at finite $G_N$. We can bound the Green's function in a microstate everywhere away from the lightcone. The bound, however, is loose, and 
passing to a thermal trace requires a separate argument. 

In section~\ref{sec:thermal} we present such an argument for the thermal state. It has a new essential element. We use joint analyticity in time and inverse temperature.
This joint analyticity only exists when the boundary theory is at finite volume. Indeed, the whole bulk cone singularity discussion is only applicable when the boundary theory is at finite volume. Using joint analyticity, we then show that for a boundary conformal field theory, at finite volume and finite $G_N$, the singular locus predicted by the bulk null rays must not exist.

Unlike the recent discussions of the bouncing geodesic, which goes all the way to the black hole singularity and reappears at complex time~\cite{HubenyLiuRangamani:2006,FidkowskiEtAl:2003,MaldacenaSimmonsDuffinZhiboedov:2015,DodelsonOoguri:2020,FestucciaLiu:2008,Bounce:KOS,Bounce:FL,Bounce:FH,Bounce:PSWambient,Bounce:CLPV,Bounce:Huang,Bounce:BGP1,Bounce:BGP2,Bounce:Valach,Bounce:CLPV2,Bounce:CV,Bounce:AJCCM,Bounce:Chak,Bounce:JR1,Bounce:JR2,Bounce:AW1,Bounce:GLS,Bounce:AH,Bounce:HLQZ,Bounce:GVV,Bounce:JK,Bounce:AEJKP,Bounce:AAL,Bounce:GMV,Bounce:DIKZ,Bounce:AW2,Bounce:HZ,Bounce:FrenkelKasner,Bounce:CGHSSquantum,Bounce:CGHSSinteriors,Bounce:DIK,Bounce:BCGHPR,Bounce:FKLW,Bounce:GM,Bounce:RGRsingularity,Bounce:KRthermal,Bounce:BMonepoint,Bounce:DK,Bounce:Singhi,Bounce:GKPdS,Bounce:BBMMPholography,Komargodski:2026Bouncing,BatraLevineShenker:2026}, here our trajectories are outside of the horizon and in fact outside of the photon sphere. For the geodesics that go all the way into the black hole singularity it is perhaps not that surprising that finite $G_N$ effects remove the singularity~\cite{Komargodski:2026Bouncing} -- after all we do expect that the black hole singularity itself disappears due to quantum gravity effects. 
But here our geodesics are completely benign and do not get too close to the horizon, let alone the singularity. So perhaps the fact that they must disappear at finite $G_N$ is surprising.  

Let us now try to understand why the singularities associated to such benign-looking null trajectories should be resolved in quantum gravity. One explanation is that the black hole metric is subject to fluctuations at finite $G_N$ and renders the trajectories in the bulk inherently fuzzy. A subset of such metric fluctuations are global fluctuations of the black hole mass itself. The thermal average takes the form 
\begin{align}
G_\beta(t,\theta)= \frac{1}{Z(\beta)}\int_{0}^{\infty} dM\, e^{S(M)-\beta M}G_M(t,\theta).
\end{align}
At large $N$, the integral over $M$ develops a saddle at the black hole mass $M(\beta)$. Expanding around this saddle gives a Gaussian distribution
\begin{align}
\frac{e^{S(M)-\beta M}}{Z(\beta)}\approx e^{-\frac{\beta^2(M-M(\beta))^2}{2C}},
\end{align}
where the positive specific heat behaves as $C\sim N^2$ at large $N$. To see how this smooths out the bulk cone, we transform (\ref{bcpower}) to the frequency domain, 
\begin{align}
G_\beta(t,\theta)\sim \int_{0}^{\infty} dM\, \int_{0}^{\infty} d\omega\, \omega^{2\Delta_\phi-\frac{5}{2}}e^{-i\omega(t-t_{\text{bulk}}(M))}e^{-\frac{\beta^2(M-M(\beta))^2}{2C}}.
\end{align}
Expanding $t_{\text{bulk}}(M)$ around $M(\beta)$ and performing the Gaussian integral, we find
\begin{align}
G_\beta(t,\theta)\sim  \int_{0}^{\infty} d\omega\, \omega^{2\Delta_\phi-\frac{5}{2}}e^{-i\omega(t-t_{\text{bulk}}(M(\beta)))}\exp\left(-\frac{(t_{\text{bulk}}'(M(\beta)))^2C}{2\beta^2}\omega^2\right).
\end{align}
Here the derivative $t_{\text{bulk}}'(M)$ scales like $N^{-2}$ at large $N$. This is the same integral that appears in the resolution of the bulk point singularity in string theory \cite{MaldacenaSimmonsDuffinZhiboedov:2015}. In the strict large $N$ limit, the Gaussian factor can be set to one, leading to a singularity. When $N$ is large but finite, the Gaussian factor resolves the integral on the scale $\Delta t\sim 1/N$.\\
\indent The discussion here has been heuristic for several reasons. First, there is no reason to expect that mass fluctuations dominate over other modes of the metric, and in particular we expect metric fluctuations to resolve the singularity even in the microcanonical example. Second, we are dealing with Planckian energies but neglecting stringy corrections. String theory is not expected to resolve the singularity by itself, but a consistent computation should include both $\alpha'$ and $G_N$ effects.\\
\indent In section \ref{3dresolution}, we make this picture concrete in a time-dependent AdS$_3$ geometry describing two colliding conical defects, first studied in \cite{Batra:2026otu}. The classical geometry admits a bulk null trajectory whose endpoints are timelike separated on the boundary, producing a bulk-cone singularity in the semiclassical correlator. By evaluating the relevant Virasoro vacuum block at finite central charge, we show that this singularity becomes a smooth peak. We then investigate the gravitational origin of this smoothing. The state preparing the defects describes a superposition of geometries with different bulk arrival times. Averaging over these fluctuations reproduces the finite-$c$ results with excellent accuracy, providing concrete evidence that quantum fluctuations of the geometry resolve the bulk-cone singularity.
\FloatBarrier
\section{Correlators in a microstate and convergence of different channels}
\label{sec:microstate}

Consider an energy eigenstate $|\mathcal{O}\rangle$ of a four-dimensional CFT on $S^3_R$. For simplicity, take the operator $\mathcal{O}$ that creates this state to be a scalar primary of dimension $\Delta_{\mathcal{O}}$, and let $\phi$ be a Hermitian scalar primary of dimension $\Delta_\phi$.

Let us recall  standard OPE convergence results~\cite{Pappadopulo:2012jk,Hogervorst:2013sma,Kravchuk:2021kwe, HartmanJainKundu:2015}.

In~$\mathbb{R}^4$, in spherical coordinates, with the  unit vectors $n,n_0\in S^3$ with $n\cdot n_0=\cos\theta$, we consider  the four-point function \begin{align}
 F_{\mathcal{O}}(r,\theta)&=\frac{\langle\mathcal{O}^{\dagger}(\infty)\phi(n_0)\phi(rn)\mathcal{O}(0)\rangle}{\langle\mathcal{O}^{\dagger}(\infty)\mathcal{O}(0)\rangle},
 \qquad 0<r<1,\\
 u=z\bar z&=r^2,\qquad v=(1-z)(1-\bar z)=1+r^2-2r\cos\theta,\\
 z&=re^{i\theta},\quad\bar z=re^{-i\theta}.
 \label{eq:microstate-crossratios}
\end{align}
We ordered the operators according to distance from the origin, since we are anticipating using radial quantization.
If one prefers the usual convention that all the operators are in the same  plane, then $ z,\bar z$
above would be the corresponding complex coordinates.

The naive OPE convergence radius is $|z|,|\bar z|<1$ since we can draw a ball containing both $\phi(n_0)$ and  $\phi(rn)$ that includes no other operator  for all  $r<1$. Alternatively, we can also use the $\mathcal{O}(0)$ and $\phi(rn)$ OPE in this domain. (These two channels overlap, which is at the foundations of the modern
conformal bootstrap.)

\begin{figure}[htbp]
\centering
\definecolor{rhoUpper}{RGB}{44,106,171}
\definecolor{rhoLower}{RGB}{202,119,36}
\definecolor{rhoContour}{RGB}{35,126,83}
\resizebox{\textwidth}{!}{%
\begin{tikzpicture}[font=\small,>=stealth,line cap=round,line join=round]
 \fill[rhoUpper!10] (-1.6,0) rectangle (2.4,1.45);
 \fill[rhoLower!12] (-1.6,-1.45) rectangle (2.4,0);
 \node[font=\bfseries] at (0.4,1.98) {$z$-plane};
 \draw[gray!70,->] (0,-1.4) -- (0,1.4);
 \draw[black!65,<-] (-1.55,0) -- (1,0);
 \draw[rhoContour,thick,dashed] (0,0) circle (1);
 \draw[white,line width=3pt] (1,0) -- (2.4,0);
 \draw[rhoUpper,very thick,->] (1,0.045) -- (2.36,0.045);
 \draw[rhoLower,very thick,->] (1,-0.045) -- (2.36,-0.045);
 \node[rhoUpper] at (1.75,0.28) {$+i0$};
 \node[rhoLower] at (1.75,-0.28) {$-i0$};
 \node at (2.59,0) {$\infty$};
 \fill (0,0) circle (1.5pt);
 \node[below left,inner sep=3pt] at (0,0) {$0$};
 \draw[fill=white] (1,0) circle (2pt);
 \node[below left,inner sep=4pt] at (1,0) {$1$};
 \node at (0.4,-1.85) {$\mathbb{C}\setminus[1,\infty)$};
 \draw[->,thick] (3.15,0.2) -- (5.25,0.2);
 \node at (4.2,0.86) {$\displaystyle\rho=\frac{1-\sqrt{1-z}}{1+\sqrt{1-z}}$};
 \node at (4.2,-0.5) {$\displaystyle z=\frac{4\rho}{(1+\rho)^2}$};
 \begin{scope}[shift={(7.3,0)}]
  \fill[rhoUpper!10] (0,0) -- (1.5,0) arc (0:180:1.5) -- cycle;
  \fill[rhoLower!12] (0,0) -- (-1.5,0) arc (180:360:1.5) -- cycle;
  \node[font=\bfseries] at (0,1.98) {$\rho$-plane};
  \draw[black!65] (-1.5,0) -- (1.5,0);
  \draw[rhoUpper,very thick] (1.5,0) arc (0:180:1.5);
  \draw[rhoLower,very thick] (1.5,0) arc (0:-180:1.5);
  \draw[rhoUpper,very thick,->] (45:1.5) arc (45:95:1.5);
  \draw[rhoLower,very thick,->] (-45:1.5) arc (-45:-95:1.5);
  \draw[rhoContour,thick,dashed]
   plot[domain=0:360,samples=241,variable=\x]
   ({1.5*(2-cos(\x)-sqrt(max(0,(2-cos(\x))^2-1)))*cos(\x)},
    {1.5*(2-cos(\x)-sqrt(max(0,(2-cos(\x))^2-1)))*sin(\x)});
  \fill (0,0) circle (1.5pt);
  \node[below left,inner sep=3pt] at (0,0) {$0$};
  \draw[fill=white] (1.5,0) circle (2pt);
  \node[right,inner sep=4pt] at (1.5,0) {$1$};
  \draw[fill=white] (-1.5,0) circle (2pt);
  \node[left,inner sep=4pt] at (-1.5,0) {$-1$};
  \node at (0,-1.85) {$|\rho|<1$};
 \end{scope}
 \node at (3.8,-2.48) {$z=0\ \mapsto\ \rho=0,\qquad z=1\ \mapsto\ \rho=1,\qquad z=\infty\ \mapsto\ \rho=-1$};
\end{tikzpicture}%
}
\caption{The conformal map~\eqref{eq:microstate-rho}.}
\label{fig:z-rho-map}
\end{figure}
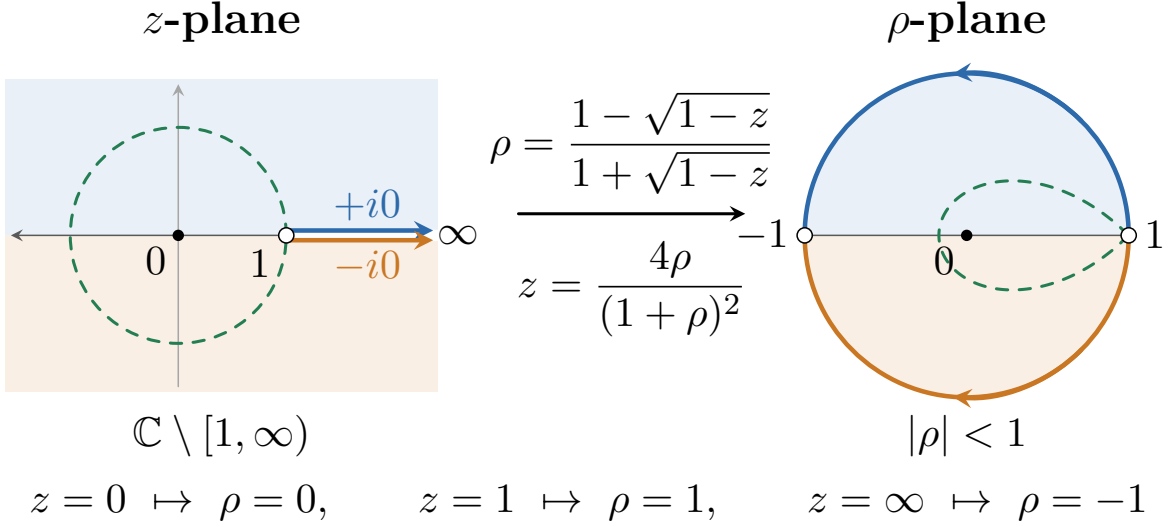

A very important result is that the convergence radius is in fact much larger. Let us focus on the $\mathcal{O}(0)$ $\phi(rn)$ OPE channel.  Introduce the $\rho$ coordinate
\begin{align}
 \rho(z)=\frac{z}{(1+\sqrt{1-z})^2}
 =\frac{1-\sqrt{1-z}}{1+\sqrt{1-z}}~.\label{eq:microstate-rho}
\end{align}
The map takes the slit plane $\mathbb{C}\setminus[1,\infty)$ to the unit disk as in Fig.~\ref{fig:z-rho-map}. Then the statement is that the OPE converges for $|\rho|<1,|\bar\rho|<1$. 

In the $\rho$ variable, the four-point function is 
\begin{align}
 F_{\mathcal{O}}(r,\theta)&=r^{-\Delta_{\mathcal{O}}-\Delta_\phi}v^{(\Delta_{\mathcal{O}}-\Delta_\phi)/2}\widehat F_{\mathcal{O}}(\rho,\bar\rho),\\
 \widehat F_{\mathcal{O}}(\rho,\bar\rho)&=\sum_{\Delta,J}p_{\Delta,J}\,
 \rho^{(\Delta+J)/2}\bar\rho^{(\Delta-J)/2},
 \qquad p_{\Delta,J}\geq0,\quad \Delta\geq |J|.
 \label{eq:microstate-radial-series}
\end{align}
The sum runs over primary and descendant states of dimension $\Delta$ and angular momentum $J$.  Euclidean OPE convergence gives for $|\rho|,|\bar\rho|\leq s<1$:
\begin{align}
 \sum_{\Delta,J}\left|p_{\Delta,J}\rho^{(\Delta+J)/2}\bar\rho^{(\Delta-J)/2}\right|
 \leq\sum_{\Delta,J}p_{\Delta,J}s^\Delta<\infty.
 \label{eq:microstate-normal-convergence}
\end{align}

To return to the cylinder, use the conformal map  
\begin{align}
 x&=e^{\tau/R}n,\qquad
 dx^2=\frac{e^{2\tau/R}}{R^2}\left(d\tau^2+R^2d\Omega_3^2\right),\\
 \phi_{\rm cyl}(\tau,n)&=R^{-\Delta_\phi}e^{\Delta_\phi\tau/R}\phi_{\rm flat}(x).
 \label{eq:microstate-cylinder-map}
\end{align}
The state insertions lie at the two ends of the cylinder, $\tau=\mp\infty$. Then on the cylinder we are computing a two-point function in a certain state
\begin{align}
 G_{\mathcal{O}}^E(\tau,\theta)&=\langle\mathcal{O}|\phi_{\rm cyl}(\tau,n_0)\phi_{\rm cyl}(0,n)|\mathcal{O}\rangle
 =R^{-2\Delta_\phi}r^{\Delta_\phi} F_{\mathcal{O}}(r,\theta),\qquad r=e^{-\tau/R}.
\end{align}
The original requirement that $r<1$ then translates to having proper time ordering on the cylinder.

For Euclidean time there is not much more to say. We can obtain $F_{\mathcal{O}}(r,\theta)$ for all $r<1$ from our absolutely convergent expansion in terms of $\rho,\bar \rho$. Actually the advantage of the $\rho$ coordinate is not yet apparent, as we could have obtained a convergent expansion using the $z$ coordinate  too. 

The advantage of the $\rho$ coordinate becomes evident when we continue to Minkowski time. To obtain the time-ordered Minkowski correlator, we  take $t>0$ and set $\tau=\epsilon+it$, with $\epsilon>0$. After this continuation,  $z=e^{-\epsilon/R-i(t/R-\theta)}$ and $\bar z=e^{-\epsilon/R-i(t/R+\theta)}$ are no longer complex conjugates.

For $\epsilon>0$, both $z$ and $\bar z$ lie inside the unit circle, which they approach as $\epsilon\to0^+$. If we did not have the $\rho$ coordinate we would not be able to say much, since the Minkowski limit of 
$\epsilon\to0^+$ would bring us to the boundary of the $z,\bar z$ convergence domain where the answer could be singular, a priori. 

However, every point on $|z|=|\bar z|=1$ except for  
$z=1$ and $\bar z=1$  maps to the {\it interior} of the $\rho$-disk. 
Thus all possible Minkowski configurations  have a complex neighborhood within the convergence domain and must lead to regular Green functions. Therefore~\eqref{eq:microstate-normal-convergence} establishes  analyticity of the  correlator in space and time.

The only possible exceptions are exactly when we land on $z=1$ or $\bar z=1$, where we have 
\begin{align}
  t/R=2\pi k\pm\theta,\qquad k\in\mathbb{Z}.
 \label{eq:microstate-null-locus}
\end{align}
 These are points joined by null geodesics on the boundary, including those that wind around a great circle. The exact microstate two-point function can therefore be singular only at null separation.

Moreover, since $p_{\Delta,J}\geq0$, taking absolute values term by term in~\eqref{eq:microstate-radial-series} bounds the real-time correlator by the Euclidean correlator in the same state at $\theta=0$ and Euclidean time $\tau_q=-R\log\frac{4q}{(1+q)^2}>0$, where $q=\max\{|\rho|,|\bar\rho|\}<1$:
\begin{align}
|G_{\mathcal{O}}^E(\epsilon+it,\theta)|\leq B^{\Delta_{\mathcal{O}}}\,G_{\mathcal{O}}^E(\tau_q,0),\qquad
B=\frac{q}{\sqrt{|\rho\bar\rho|}}\,\frac{|1-\rho|\,|1-\bar\rho|}{(1-q)^2}.
\label{eq:microstate-explicit-bound}
\end{align}
The right-hand side is finite, as $\epsilon\to0^+$, everywhere away from~\eqref{eq:microstate-null-locus}. For black hole states we have $\Delta_{\mathcal{O}}\sim N^2$ and hence this bound is parametrically exponential in $N^2$, which means it is not an effective bound. 

It is instructive to repeat the analysis in the light--light channel $\phi(rn)\times\phi(n_0)$, which is the natural channel in the $1/N$ expansion. Later we will consider an example of how we can utilize this channel. The adapted radial coordinate is $\varrho(z)=(1-\sqrt z)/(1+\sqrt z)$, which maps $\mathbb{C}\setminus(-\infty,0]$ to the unit disk, with the OPE point $z=1$ at the center and the heavy insertions $z=0,\infty$ at $\varrho=\pm1$. The correlator is a power series in $\varrho,\bar\varrho$ with coefficients proportional to $C_{\phi\phi X}C_{\mathcal{O}^\dagger\mathcal{O}X}$, which are no longer sign-definite, though by Cauchy--Schwarz the series still converges absolutely for $|\varrho|,|\bar\varrho|<1$. In real time, as $\epsilon\to0^+$,
\begin{align}
 \varrho\to i\tan\frac{t/R-\theta}{4}~,\qquad \bar\varrho\to i\tan\frac{t/R+\theta}{4}~,
 \label{eq:microstate-light-light}
\end{align}
so this channel establishes analyticity away from the first light cone $t/R=\theta$ (where $\varrho=0$) only in the window $t/R+\theta<\pi$.

This is a weaker result than the one obtained in the heavy-light channel, and as far as we can see it cannot be improved. At $t/R+\theta=\pi$ the variable $\bar z$ reaches the cut and then continues to the second sheet around the heavy operator at $\bar z=0$, where $|\bar\varrho|>1$. Since $\bar\varrho=\pm1$ are genuine branch points, the radius of convergence is exactly one and the light--light expansion diverges there, even though we have just shown that the correlator itself is analytic. The bulk-cone locus of Fig.~\ref{fig:bulk-cone-temperatures} lies at $t/R+\theta>\pi$ and is therefore entirely outside the reach of the light--light OPE. 

This is as it should be: it is the divergence of this light-light OPE on the second sheet that allows the $1/N$ expansion to develop a bulk-cone singularity at $G_N=0$, while it is the positivity of the heavy--light channel that excludes it at finite $G_N$. 
Indeed, when we perform the $1/N$ expansion in the light-light channel, if we take the large $N$ limit first, we do not obtain a consistent light-heavy OPE, which allows for the bulk-cone singularity. 

The fact that the Green's function is finite on the cylinder away from the light-cone does not immediately imply that the thermal cylinder Green's function is finite away from the light cone (although it is very suggestive). 

To prove that the thermal cylinder Green's function is finite away from the light cone we will need a new idea.
And in fact we will not be able to prove this result in full generality, but we will be able to prove a sufficiently strong result that rules out the bulk cone singularity.

\section{Finite temperature and finite volume}
\label{sec:thermal}

Consider a finite temperature field theory on a compact manifold. The Wightman function for a Hermitian operator $\mathcal{O}$ takes the form
\begin{align}\label{traceconv}
G_{\beta}(t,x,y)&=\frac{1}{Z(\beta)}\text{Tr}(e^{-\beta H}\mathcal{O}(t,x)\mathcal{O}(0,y))\notag\\
&=\frac{1}{Z(\beta)}\sum_{mn}e^{-(\beta-it)E_n-itE_m}\mathcal{O}_{nm}(x)\mathcal{O}_{mn}(y),
\end{align}
with the matrix elements 
\begin{align}
\mathcal{O}_{mn}(x)=\langle  m|\mathcal{O}(x)|n\rangle.
\end{align}
Here $Z(\beta)=\text{Tr}(e^{-\beta H})$ is the thermal partition function. For a theory in $d$ spacetime dimensions, the high-energy behavior of the entropy is
\begin{align}\label{dos}
S(E)\propto E^{(d-1)/d},\hspace{10 mm}E\to \infty.
\end{align}
This implies a subexponential density of states, so that $Z(\beta)$ admits an analytic continuation to the half-plane $\operatorname{Re} \beta>0$, 
\begin{align}
Z(\beta)=\int\,  dE\, e^{S(E)-\beta E}<\infty.
\end{align}
Note that finite volume is crucial here, since entropy scales extensively with volume.\\
\indent Let us now complexify the two variables $t$ and $\beta$. 
The function $Z(\beta)G_\beta(t,x,y)$ can be bounded using the Cauchy-Schwarz inequality, 
\begin{align}
|Z(\beta)G_\beta(t,x,y)|&\le\sum_{mn}e^{-(\operatorname{Re}(\beta)+\operatorname{Im}(t)) E_n+\operatorname{Im}(t)E_m}|\mathcal{O}_{mn}(x)\mathcal{O}_{nm}(y)|\notag\\
&\le |Z(\operatorname{Re} \beta)|\sqrt{|G_{\operatorname{Re} \beta}(i\operatorname{Im} t,x,x)G_{\operatorname{Re} \beta}(i\operatorname{Im} t,y,y)|}.
\end{align}
Here $G_{\operatorname{Re} \beta}(i\operatorname{Im} t,x,x)$ is the coincident-point Euclidean correlator $G_E(\tau,x,x)$, with $\tau=-\operatorname{Im} t$. It follows that $Z(\beta)G_\beta(t,x,y)$ is finite in the region
\begin{align}\label{analyticity}
\operatorname{Re} \beta>0,\hspace{10 mm}0>\operatorname{Im} t>-\operatorname{Re} \beta.
\end{align}
 The normalized correlator $G_\beta(t,x,y)$ is then analytic in the same region, as long as one stays away from the complex zeroes of $Z(\beta)$. Note that it is essential that $\operatorname{Im} t$ is not strictly zero, since there can be light-cone singularities on the real $t$ axis.
\\
\indent We aim to show that (\ref{analyticity}) rules out any spurious singularities away from the light cone. Consider a putative bulk-cone singularity of the form 
\begin{align}\label{bulkcone}
Z(\beta)G_\beta(t,x,y)&\sim \frac{A(\beta)}{(t-T(\beta))^{\gamma}}.
\end{align}
Here we leave the $x$ and $y$ dependence of the functions $A$ and $T$ implicit. The function $T(\beta)$ is real and locally analytic for physical $\beta$. Assume that $T(\beta)$ is not constant. When $\beta$ is complexified, the open mapping theorem ensures that an open neighborhood around $\beta_0$ maps to an open neighborhood around $T(\beta_0)$. This image necessarily overlaps with $\operatorname{Im} t< 0$, contradicting the analyticity property (\ref{analyticity}). This implies that $T(\beta)$ is independent of $\beta$, so only light-cone singularities are allowed\footnote{In principle we could have some other singularity away from the light-cone with constant $T(\beta)$, as long as $A(\beta)$ vanishes at zero temperature. We are not aware of any examples of such singularities.}.
\begin{figure}
    \centering
    \includegraphics[width=0.5\linewidth]{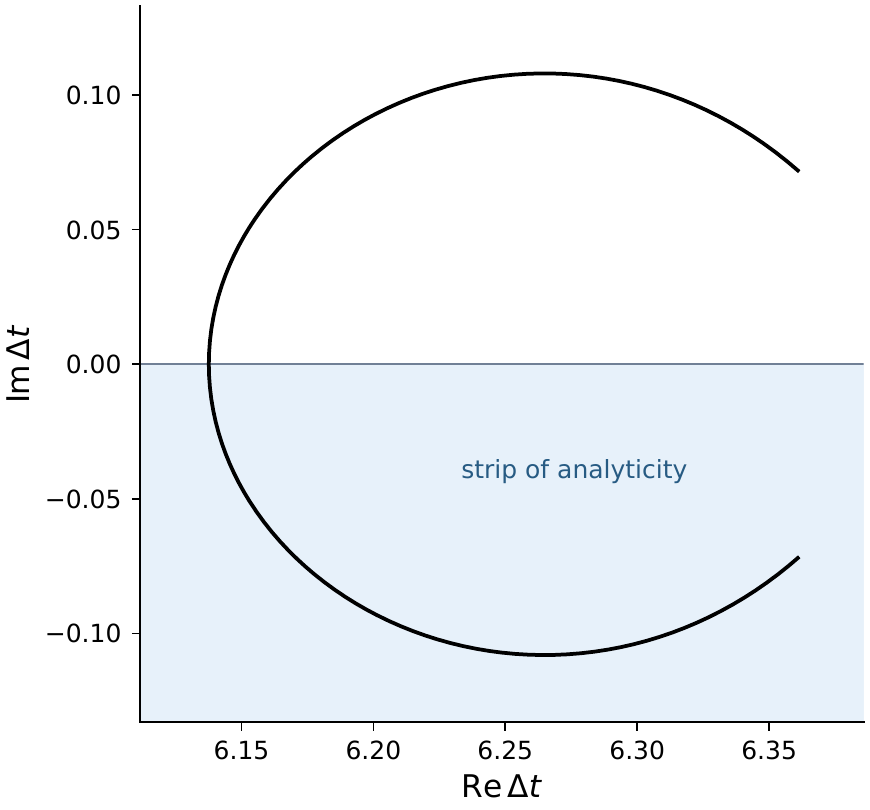}
    \caption{The location of the bulk-cone singularity for fixed $\Theta$. We take $\beta=2\pi/3+is$ and vary $s$ from $-1$ to 1. }
    \label{fig:placeholder}
\end{figure}
\\\indent It is instructive to return to the example presented in the introduction. The integrals in (\ref{eq:intro-return-locus}) can be performed explicitly,
\begin{align}
t_{\text{bulk}}&=\frac{2}{b(1+2y_h^2)\sqrt{1-q}}\left(y_h^2\Pi\left(\frac{q}{1+y_h^2},\frac{q}{1-q}\right)+(1+y_h^2)\Pi\left(-\frac{q}{y_h^2},\frac{q}{1-q}\right)\right),\\
\Theta&=\frac{2}{\sqrt{1-q}}K\left(\frac{q}{1-q}\right),
\end{align}
where we have defined 
\begin{align}
q=\frac{y_h^2(1+y_h^2)}{y_*^2}.
\end{align}
Fixing the elapsed angle $\Theta$ to some real value with $\Theta>\pi$ singles out a unique value of the parameter $q$.                                                             \\
\indent For fixed $q$, the parametric relation $(\beta,t_{\text{bulk}}(\beta))$ traces out a two-dimensional Riemann surface as $\beta$ is varied in the complex plane. Due to the elliptic $\Pi$ functions, this surface is infinitely sheeted. Curiously, the surface extends even to $\operatorname{Re} \beta<0$, where there is no guarantee of analyticity. In figure \ref{fig:placeholder} we plot the trajectory of $t_{\text{bulk}}(\beta)$ as a function of $\operatorname{Im} \beta$ for fixed $\operatorname{Re} \beta$. Consistently with the previous argument, this curve enters the disallowed strip of analyticity.

\section{Bulk-cone resolution in three dimensions}\label{3dresolution}

We study a bulk cone singularity produced by two colliding defects in AdS$_3$ \cite{Batra:2026otu}, identify its boundary counterpart, and show how finite $G_N$ resolves it. We consider three dimensions because it offers analytical control over the  gravitational degrees of freedom, the boundary gravitons. Our example is motivated by bulk cone singularities in Vaidya-like states~\cite{Leung:2024}, whose collapsing shells can be modeled in the CFT by many light operator insertions~\cite{Anous:2016,Anous:2017}. Related studies of scalar propagation and boundary correlators in static conical-defect backgrounds appear in~\cite{BerensteinGrabovskyLi:2023,Grabovsky:2024}.

Away from the particle worldlines, three-dimensional vacuum Einstein gravity with negative cosmological constant is locally AdS$_3$. We construct the spacetime by excising wedges from an AdS covering region and identifying their faces ~\cite{banados,Matschull:1998rv}. With the AdS radius set to one, we use global coordinates
\begin{align}
	ds^2 = -{\cosh^2 \rho} \, dt^2 + {d\rho^2} + \sinh^2 \rho \,d\varphi^2,
\end{align}
and choose the particles to approach the origin along $\varphi=\pm\pi/2$. For each particle, we begin with an auxiliary static defect: an angular wedge is removed, its edge lies at $\rho'=0$, and its two faces are identified. An AdS boost maps this edge to a moving worldline and transforms both wedge faces and their identification. We choose a separate boost for each defect, excise both resulting wedges from a common AdS covering region, and identify each pair of faces~\cite{Lindgren:2016wtw}. These independent boosts specify the relative configuration of the two particles; a single global boost could not separate coincident worldlines.

For example, equal and opposite rapidities $\pm\chi$, with $\chi>0$, give the incoming trajectories
\begin{equation*}
    \varphi_\pm=\pm\frac{\pi}{2},\qquad
    \tanh\rho_\pm(t)=\tanh\chi\cos t,
    \qquad 0\leq t<\frac{\pi}{2}.
\end{equation*}
The particles therefore start on opposite sides at $\rho_\pm(0)=\chi$ and meet at the origin at $t=\pi/2$. Each boost preserves the corresponding deficit angle because it is an AdS isometry, but changes the locations of the identified wedge faces on the boundary.

To determine these boundary identifications, let $(t',\rho',\varphi')$ denote the auxiliary rest-frame coordinates of one defect and $(t,\rho,\varphi)$ the common covering coordinates. Each defect has its own primed chart. The near-boundary limit of an embedding-space boost of rapidity $\chi$ gives
\begin{align}
	\Lambda \sin t  &=  \sin t', \label{eq:boost1}\\
	\Lambda  \cos t &=  \cosh \chi \cos t'  + \sinh \chi \sin \varphi', \label{eq:boost2} \\
	\Lambda  \cos \varphi  &=   \cos \varphi' , \label{eq:boost3}\\
	\Lambda \sin \varphi  &=  \cosh \chi \sin \varphi'+ \sinh \chi \cos t' . \label{eq:boost4}
\end{align}
Here $\chi$ is the rapidity and $\Lambda=e^{\rho-\rho'}$ is the radial rescaling in the near-boundary limit. We orient each excised wedge away from the other particle and toward the nearer boundary. For the defect on the $\varphi=+\pi/2$ axis, the rest-frame faces at $\varphi'=\varphi'_+$ and $\varphi'=\pi-\varphi'_+$  are identified. Thus $\varphi'_+$ is a constant face angle in the auxiliary rest frame; the prime denotes that frame, not a derivative. The removed wedge has opening angle $\pi-2\varphi'_+$, equal to the deficit angle. The boost changes the boundary locations of its faces as illustrated below:
\begin{align}
	\includegraphics[scale=0.5, valign = c]{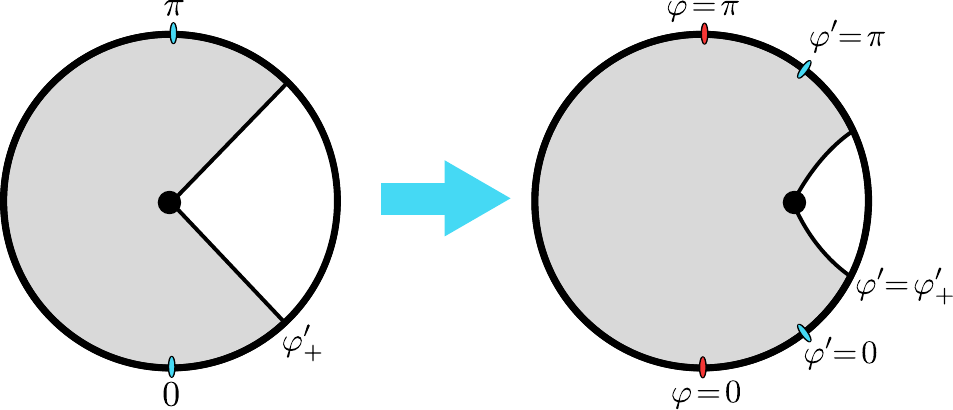}
\end{align}
Writing these boundary locations as $\varphi=\Phi_+(t)$ and $\varphi=\pi-\Phi_+(t)$, we obtain
\begin{align}
	\cos t \cos \varphi'_+  \sinh \chi_+ = \sin \Phi_+(t) \cos \varphi'_+ \cosh \chi_+ - \cos \Phi_+(t)  \sin \varphi_+', \label{eq:id_path}
\end{align}
from \crefrange{eq:boost1}{eq:boost4}, with $\chi_+$ the rapidity of this defect. The other defect is constructed in the same way with its own rest-frame face angle and boost. After both wedges have been excised and each pair of faces identified, the configuration is
\begin{align}
	\includegraphics[scale=0.5, valign = c]{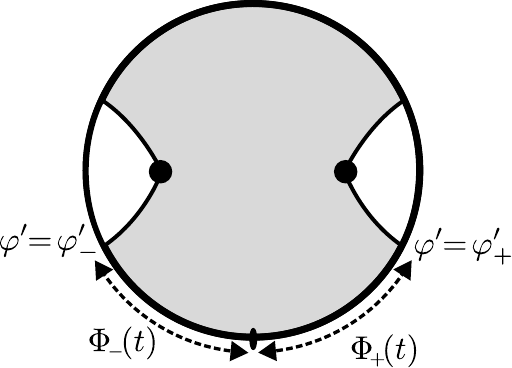}
\end{align}
The quotient locations from the origin solve to
\begin{align}
	\Phi_i(t) = \arctan\left[ \tan \varphi'_i \over \cosh \chi_i\right] + \arcsin \left[ {\cos \varphi_i' \sinh \chi_i \cos t \over \sqrt{1 + \cos^2 \varphi_i' \sinh^2 \chi_i} }\right],
\end{align}
where the index $i = \pm$ refers to the two defects near $\varphi = \pm \pi/2$ and identifies $\varphi[\varphi'_\pm] = \Phi_\pm \sim \varphi[\pm \pi - \varphi'_\pm] = \pm\pi - \Phi_\pm(t)$.\footnote{We are working with the branch $\varphi \in [-\pi, \pi]$.}

It's not hard to see that these identifications lead to bulk cone singularities. We give a qualitative argument here and perform a detailed check later in this section. Compare the covering coordinate arrival times for bulk and boundary null geodesics that begin at $\varphi = 0$ and end at $\varphi = \pi$. For a geodesic traveling through the bulk, the elapsed covering time is given by
\begin{align}
	\Delta t = 2\int_0^\infty{d\rho \over \cosh \rho } = \pi.
\end{align}
Without boundary quotients, these two points are separated by $\Delta \varphi = \pi = \Delta t$ and therefore are on the boundary lightcone. In pure AdS, bulk and boundary null rays therefore arrive simultaneously~\cite{GaoWald:2000}. With quotients, boundary null paths experience a discontinuous spacelike jump in the  $\varphi$ coordinate; a null path propagating along $\varphi = [0,\pi]$ looks like 
\begin{align}
	\includegraphics[scale=0.7, valign = c]{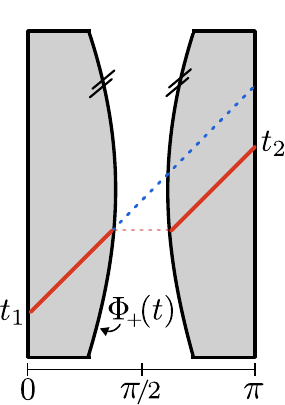}
\end{align}
When a boundary null geodesic (red) hits $\varphi = \Phi_+(t)$ it gets transported (dotted red) to $\varphi = \pi -\Phi_+(t)$, after which it continues to  $\varphi= \pi$. This implies the two endpoints $(t_1, 0)$ and $(t_2, \pi)$ are spacelike instead of null separated in the $(t,\varphi)$ coordinates. Therefore, the arrival time, in the coordinate $t$, is less than $\pi$. We emphasize that the discontinuity, and apparent acausality, of this path is an artifact of working in the fiducial covering  coordinate frame. In the physical boundary coordinates, which we now determine, it is a genuine boundary null trajectory. We now turn to a quantitative check of these claims.

{\bf \flushleft Determining the holographic stress tensor:} The map to physical coordinates is determined by working in Fefferman--Graham (FG) gauge~\cite{FeffermanGraham:2012}, finding the holographic stress tensor, and identifying the covering coordinates with the coordinates that uniformize away the  stress tensor: we find the conformal transformation that sets it to zero. In Euclidean asymptotically AdS spacetime, we solve Einstein's equation in the gauge
\begin{align}
	ds^2 = d\rho^2 + \cosh^2 \rho \, e^{\phi}\,|dw|^2,
\end{align}
where the only nontrivial equation requires the two-dimensional metric $e^{\phi}|dw|^2$ to have constant negative curvature. In  vacuum, this gauge describes the AdS$_2$ slicing of AdS$_3$. In the presence of the defects, it requires $\phi$, the dilaton, to satisfy the Liouville equation
\begin{align}
	\partial \bar{\partial} \phi = {e^{\phi} \over 2} -  \pi \sum_{i} \gamma_i \delta(w - w_i),
\end{align}
where $w_i$ are the defect locations and $2\pi \gamma_i$ is the deficit angle.  This equation does not have a known closed form solution for two arbitrary defects. However, when the defects are far apart from each other, an approximate solution is
\begin{equation}
\begin{aligned}
    \phi &=    \ln\left[ {4 \over (1 - |w|^2)^2}\right] + \sum_{i = \pm}\ln\left[ (1 - \gamma_i)^2(1 - |v_i(w)|^2)^2\over |v_i(w)|^{2 \gamma_i}(1 - |v_i(w)|^{2(1 - \gamma_i)})^2 \right]\\
    &\equiv \phi_0 + \phi_+ + \phi_-,
\end{aligned}
\label{eq:approx_solution}
\end{equation}
where $v_i(w) = (w - w_i)/(1 - \bar{w}_i w)$ is the M\"obius transformation that boosts $w = w_i$ to the origin. It is not hard to check that this solution satisfies the necessary boundary conditions at the defect and the boundary
\begin{equation}
\begin{aligned}
    \phi &\sim -2 \gamma_i\ln |w - w_i|,\quad w\sim w_i, \\
    &\sim -2\ln \left( 1 - |w| \right), \quad |w| \to 1.
\end{aligned}
\end{equation}
The second condition imposes AdS$_2$ asymptotics. The error from plugging this solution into the differential equation is $(1 - e^{-\phi_+})(1 - e^{-\phi_-}) \sim e^{-2 D(w_+,w_-)}$ where $D(w_+,w_-)$ is the geodesic distance between $w_\pm$ in AdS units. Hence, the approximate solution \eqref{eq:approx_solution} is justified for large relative boosts between the defects. 

The asymptotic metric can be put in FG form,
\begin{align}
    d\rho^2 + \cosh^2 \rho \, e^{\phi}|dw|^2 \approx {dy^2 + |dz|^2\over y^2} \quad \mathrm{as} \ \rho,y^{-1} \to \infty,
\end{align}
through the coordinate transformation
\begin{gather}
	\rho = -\ln {ye^{\phi/2}  \over 2} + \cdots, \quad w = z - {y^2 \over 2} \bar{\partial} \phi + \cdots, \quad \bar{w} = \bar{z} - {y^2 \over 2} {\partial} \phi + \cdots.
\end{gather}
This coordinate transformation is trivial at leading order because of the large boost that concentrates the defect effects near the two boundary points $w_{\pm}$. Nevertheless, it still leads to non-trivial physics, as we will see below. The metric in these coordinates can be cast in  Ba\~nados form \cite{banados}
\begin{align}
    ds^2  = {dy^2 + |dz|^2 \over y^2 } - {6 \over c}T dz^2 - {6 \over c}\bar{T} d\bar{z}^2 + {36 \over c^2}y^2 T \bar{T}|dz|^2,
\end{align}
where $T = {c \over 6} \left( \partial^2 \phi/2 - (\partial  \phi)^2/4 \right)$, the stress energy tensor, is
\begin{align}
	T &= \sum_{i = R,L} {h_i(1 - |z_i|^2)^2\over (z - z_i)^2(1 - \bar{z}_i z)^2} \\
	&= \sum_{i = R,L}h_i \left({1 \over (z - z_i)^2} + \frac{2\bar{z}_i}{(1 - |z_i|^2)(z - z_i)} + {\bar{z}_i^2 \over (1 - \bar{z}_i z )^2} + \frac{2\bar{z}_i^2}{(1 - |z_i|^2)(1 - \bar{z}_i z)} \right),\notag
\end{align}
and similarly for the complex conjugate. Notice that the stress tensor picks up singularities at two locations on the boundary per defect; those are the boundary endpoints of the bulk defect trajectory which, in Euclidean AdS, actually reach the boundary. The absence of terms coupling the two defects is another consequence of the large boost, consistent with a pairwise same-defect OPE limit on the boundary.

{\bf \flushleft Uniformizing the embedding AdS$_3$ coordinate:} The next step is to find the uniformizing coordinate.  Since the bulk cone problem is a Lorentzian problem, we  analyze the Lorentzian continuation of the solution above. Consider first transforming the plane coordinate $z$ to the cylinder using $x = \log z$ in which the defects are at $x = i \pi/2\pm\epsilon_R, -i \pi/2 \pm \epsilon_L$, remaining on the boundary Euclidean section.\footnote{We will henceforth use  $R,L$ labels to refer to the two bulk defects and their pair of boundary counterparts near $x = 0,i\pi$, respectively.} For a single defect, $\epsilon$ is related to the boost through $\sinh \epsilon = 1/\sinh \chi$. The Lorentzian continuation is then implemented by shifting $x \to x + i \hat{t}$. Working in the limit where $\epsilon \to 0$ and holding $p \equiv 12 \pi h/\epsilon c$ fixed, the Lorentzian stress tensor simplifies to
\begin{equation}
\begin{aligned}
	T_{uu} &=- {c \over 24} +  \sum_{i = R,L} {h_i\sinh^2 \epsilon_i\over (\cosh \epsilon_i - \cosh (x_i + i \hat{t} ))^2}  \\
	&= -{c \over 24}  +  {c{p_R^+} \over 6} \delta(\hat{t} + \hat{\varphi}- \pi/2) +  {c p_L^+\over 6} \delta(\hat{t} + \hat{\varphi} + \pi/2),
\end{aligned}
\end{equation}
where $u \equiv \hat{t} + \hat{\varphi}$. The uniformizing map is the transformation to the frame where the shift of the stress tensor due to the Weyl anomaly sets it to zero. It satisfies
\begin{align}
	T_{uu} = -{c \over 12} \{w(u),u\},
\end{align}
where $w$ is the covering coordinate on the plane. Solutions of this Schwarzian equation can be written as $w(u) = {\psi_1(u) \over \psi_2(u)}$ where  $\psi_{1,2}$ are two linearly independent solutions to the second-order differential equation
\begin{align}
	\psi_k''(u) - {6 \over c} T_{uu} \psi_k = 0 \implies w(u) = {\psi_1(u) \over \psi_2(u)}.
\end{align}
We will work with the cylinder covering coordinate using $w(u) = \tan {g(u) \over2}$. Working in the shockwave limit, the general solution is a M\"obius transformation of $\tan {u \over 2}$ away from the delta-function stress energy pulses at $u =  \pm{\pi \over 2}$.

The maps $g(u),g'(u)$ are continuous everywhere but $g''(u)$ is discontinuous across the boundary stress tensor shockwaves. It'll be instructive to work out an example. Consider a solution where $g(u) = u$ in the region $-\pi/2<u < \pi/2$. We can write this solution as 
\begin{align}
	\tan {g_0(u) \over 2} &= { B_1\cdot T  \over  B_2\cdot T},
\end{align}
where
\begin{gather}
	B_1 = \begin{pmatrix}
		1 &  0
	\end{pmatrix}, \hspace{10 mm}B_2 = \begin{pmatrix}
		0 & 1
	\end{pmatrix}, \hspace{10 mm}T = \begin{pmatrix}
		\tan {u \over 2} \\ 1 
	\end{pmatrix}.
	 \end{gather}
We can think of $B_{1,2}$ as picking out an SL$_2$ gauge. Upon crossing a shockwave at $u = \pi/2$ from below or $u = -\pi/2$ from above, the solution transforms to
\begin{gather}
\tan {g_R(u) \over 2}  = { B_1\cdot P_R^+ \cdot T  \over  B_2\cdot P_R^+ \cdot  T} \quad \mathrm{or} \quad \tan {g_L(u) \over 2}  = { B_1\cdot P_L^+ \cdot T  \over  B_2\cdot P_L^+ \cdot  T}, 
\end{gather}
where 
\begin{align}
	P_{R}^+ = \begin{pmatrix}
		1 + p_{R}^+ & -p_{R}^+ \\ p_{R}^+ & 1 - p_{R}^+
	\end{pmatrix}, P_{L}^+ = \begin{pmatrix}
		1 + p_{L}^+ & p_{L}^+ \\ -p_{L}^+ & 1 - p_{L}^+
	\end{pmatrix}.
\end{align}
These rules can be generalized to crossing any number of shockwaves. Pictorially, this refers to
\begin{align}
    \includegraphics[scale=0.45, valign = c]{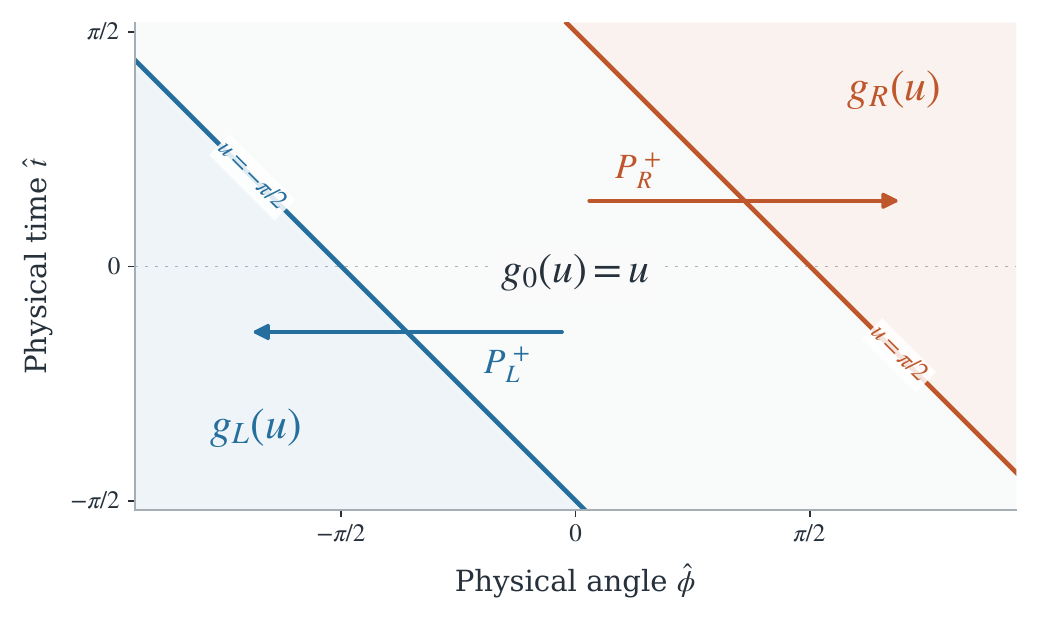}. 
\end{align}

{\bf \flushleft Evolution of the quotient:} 
With the mapping at hand, we can determine the time evolution of the identifications in the covering coordinates. We extend the above construction to the other null direction in both coordinates: $f \equiv t - \varphi$ in the covering coordinate and $v = \hat{t} - \hat{\varphi}$ in the physical coordinate. We pick a gauge where the mapping is trivial $f(v) = v$ and $g(u) = u$ in the region $-\pi/2 < u,v < \pi/2$ between the stress energy shockwaves. The shockwave limit actually allows us to say more than that. Since the bulk geometry of the $\hat{t} = t = 0$ slice is almost exactly pure AdS$_3$ away from the defect insertions near the boundary at $\hat{\varphi} = \varphi = \pm \pi/2$, the coordinates in the causal diamond  $-\pi/2 < u,-v < \pi/2$ can be extended uniquely through the bulk to those in the region $\pi/2 < u,-v < 3\pi/2$. Therefore, the vacuum gauge $f(v) = v$ and $g(u) = u$ can be simultaneously imposed in both regions.

\begin{figure}%
    \centering
    {{\includegraphics[width=8cm]{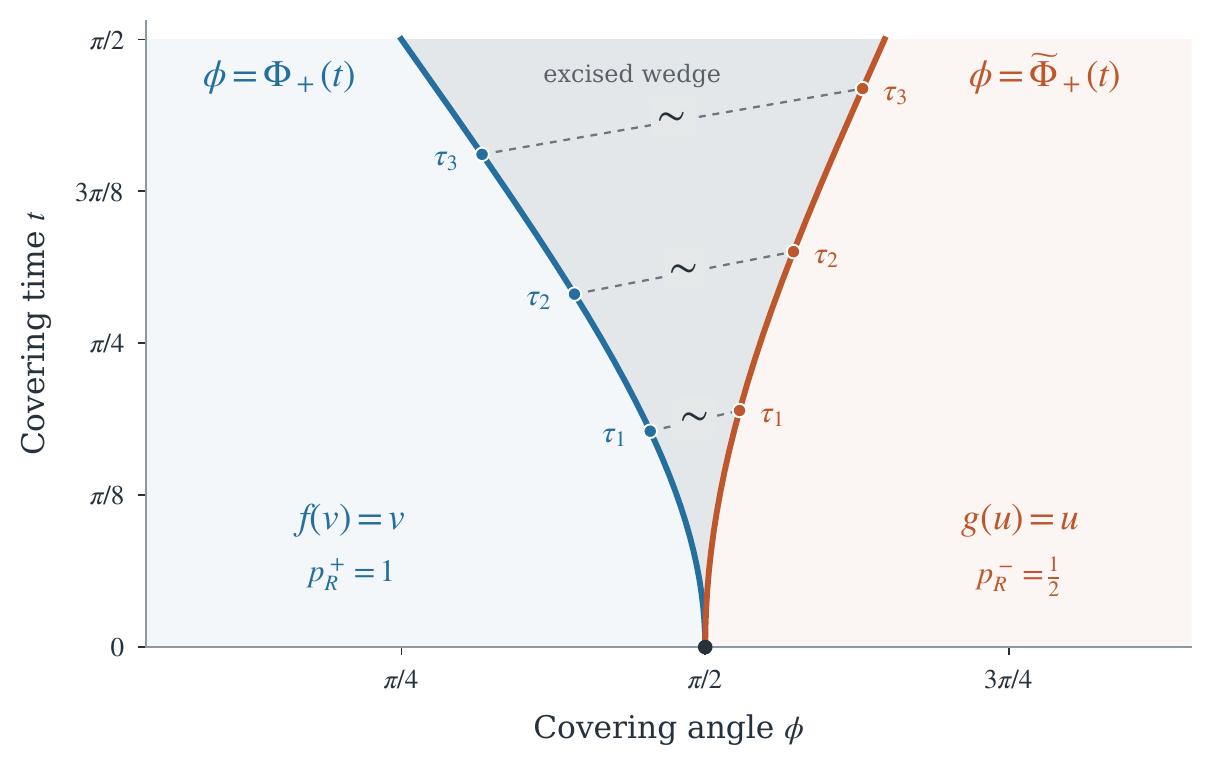} }}%
    \caption{The physical line $(\hat{t}, \hat{\varphi}) = (\tau, \pi/2)$ is mapped to two different lines in the covering coordinate, each constructed from a different coordinate map starting from different patches.}%
    \label{fig:quotient_evolution}%
\end{figure}

Off the $t = 0$ slice, the covering and physical coordinates deviate from one another, the latter describing less and less of the AdS$_3$ spacetime as the defects fall into the bulk and the identified wedge region grows. We can obtain the shape of where the wedge meets the boundary by extending the trivial solutions on either side of each defect insertion to their future. The wedge is then obtained by selecting a physical timelike path covered by the extension of both coordinates and finding its images under both maps. A convenient choice is $(\hat{t}, \hat{\varphi}) = (\tau ,\pi / 2)$ for $\tau > 0$. In this region $-v < \pi/2$ while $u > \pi/2$, and therefore we can maintain $f(v) = v$ while $g(u)$ is modified by the shock to satisfy
\begin{align}
	\tan {g(u) \over 2} = {(1 + p_R^+)\tan {u \over 2} - p_R^+ \over p_R^+ \tan {u \over 2} + (1 - p_R^+)}.
\end{align}
Denote by $\Phi_+(t)$ the location of the first boundary of the wedge from the origin, as we did in \eqref{eq:id_path}. Combining $f = t - \Phi_+(t) = \tau - \pi/2$ with the equation for $g$ we arrive at
\begin{align}
	p_R^+ \cos t  = p_R^+ \sin \Phi_+(t) - \cos \Phi_+(t),
\end{align}
which agrees with the path in the covering coordinates we found earlier \eqref{eq:id_path} but in the infinite boost limit where $\cos \varphi'_+ \cosh \chi_+ = p_R^+$ is held fixed as $\varphi'_+ \to \pi/2$ and $\chi_+ \to \infty$. The path of the other boundary of the wedge is obtained by starting with the trivial solution in $\pi/2 < u,-v < 3\pi/2$ and crossing into $-v < \pi/2$. In this case the $f(v)$ map evolves to
\begin{align}
	\tan {f(v) \over 2} = {(1 - p_R^-)\tan {v \over 2} - p_R^- \over p_R^- \tan {v \over 2} + (1 + p_R^-)},
\end{align}
leading to the path
\begin{align}
	p_R^- \cos t  = p_R^- \sin ( \pi - \tilde{\Phi}_+(t)) - \cos( \pi -  \tilde{\Phi}_+(t)).
\end{align}

{\bf \flushleft An example of a bulk cone singularity:} We now derive a bulk cone singularity. For simplicity, we assume scalar defects with equal weights. The strategy is to determine the physical location of the bulk cone singularity, which we label as $(\tilde{t}, \tilde{\varphi})$, and then show that it takes less time for null signals on the boundary to reach $\tilde{\varphi}$: we want to show that $\tilde{t} - \tilde{\varphi} - t_1 > 0$ where $(t_1,0)$ are the launch coordinates.

Consider a null geodesic that launches from $(t,\varphi) = (\hat{t}, \hat{\varphi}) = (-\pi/4,0)$. Through the bulk, this geodesic  travels through pure global AdS$_3$ and  arrives at the covering coordinates $(3\pi/4, \pi)$. The physical coordinates can then be obtained by determining in which patch this singularity resides, and using the corresponding map between physical and covering coordinates. Switching to null coordinates, $(u,v) = (\hat{t} + \hat{\varphi}, \hat{t} - \hat{\varphi})$, we have to pinpoint $(\tilde{t} + \tilde{\varphi}, \tilde{t} - \tilde{\varphi})$.  Starting with the trivial maps defined in the region $ \pi/2<g,-f<3\pi/2$, we ask how many shocks are crossed as $g,f$ are moved to the future. As we do that, we have to determine the branch of the coordinate map i.e. how many times $u,v$ wind as $g,f$ are increased. The formula we obtain for the arrival location is
\begin{align}
	\tilde{u} &\equiv  (j_+ + 1/2)\pi -2 \arctan \left[ \left(p_L^+ \over {p_R^+} \right)^{(-1)^{j_{+}}/2} {\sin \left( (j_+ - 1)\theta_+ + \delta_+[t_1] \right)\over \sin \left( j_+\theta_+ + \delta_+[t_1] \right)} \right],\\
	\tilde{v} &\equiv  (j_- - 3/2)\pi -2 \arctan \left[ \left(p_R^- \over {p_L^-} \right)^{(-1)^{j_{-}}/2} {\sin \left((j_- - 1)\theta_-+ \delta_-[t_1] \right)\over \sin \left( j_-\theta_- + \delta_-[t_1] \right) }\right],
\end{align} 
where $j_i$ counts the number of windings of the physical coordinates, $\delta_i \in [0,\pi]$ is a phase offset and satisfies $\theta_i < \delta_i < \pi$, where $\theta_i \in(0,\pi)$ is a measure of the strength of the shockwaves,
\begin{gather}
	j_i[s] \equiv \left\lceil { \pi - \delta_i[s] \over\theta_i}\right\rceil, \quad  \cos \theta_i = \sqrt{p_R^i p_L^i},
 \\  \delta_i[s] \equiv {\pi \over 2} +   \arctan\left[ {\left( p_R^i \over p_L^i\right)^{{{\mathrm{sgn}[i]} \over 2}} } \tan\left(\frac{\pi}{4}-\frac{s}{2}\right) -  \sqrt{p_R^i p_L^i} \over \sin \theta_i \right].
  \end{gather}
These formulas apply only for $p_L^ip_R^i < 1$, below the black hole formation threshold; otherwise they have to be continued appropriately. Consider the case $p_L^\pm = p_R^\pm = 1/2$. Evaluating these parameters gives
\begin{gather}
	j_+ = j_- = 1, \quad \theta_+ = \theta_- = \arccos[1/2] \approx 1.047, \\
	\delta_\pm = {\pi \over 2} + \arctan{1 + 2 \sqrt{2} \over \sqrt{3}} \approx 2.717, \\
	\implies \tilde{u} = {3\pi \over 2} + 2 \arctan {1 \over  \sqrt{2}},  \tilde{v} = -{\pi \over 2} + 2 \arctan {1 \over  \sqrt{2}}.
\end{gather}
We find the physical spatial arrival location is $\tilde{\varphi} = (\tilde{u} - \tilde{v})/2 = \pi$, consistent with the bulk arrival location. The  arrival time of the bulk null geodesic is $\tilde{t} = \pi/2 + 2 \arctan{1 \over  \sqrt{2}} \approx 2.80$, and so the physical boundary time lag is $\tilde{t} - \tilde{\varphi} - t_1 = 2 \arctan{1 \over  \sqrt{2}} - \pi/4 \approx 0.4456$, indicative of a bulk cone singularity.

We emphasize that the discussion above applies equally well to the large $c$ probe two-point function in the background prepared by two heavy operators. As shown in \cite{FKW}, the background computation is the same as evaluating the identity block contribution to the correlator of four heavy primary operators. The probe two-point function is then obtained by transforming to the appropriate conformal frame using the maps derived above: 
\begin{align}
   \! \! \!  &{1 \over \left( \sin {u_2 - u_1 \over 2} \right)^{2h}} = \left({\left( \cos{u_2 \over 2} \cos{u_1 \over 2} \right)^{-2}  \over \left( \tan{g(u_2) \over 2} - \tan {u_1 \over 2} \right)^2} \right)^{h} \! \! \to \left({\left(\tan {g(u_2) \over 2} \right)' \over \cos^2{u_1 \over 2} \left( \tan{g(u_2) \over 2} - \tan {u_1 \over 2} \right)^2} \right)^{h} \\
   &= {1 \over  \left( \sin {u_2 - u_1 \over 2} - 2 p_L^+ \cos {u_2 - \pi/2 \over 2}\cos {u_1 - \pi/2 \over 2} \right)^{2h} }. \label{eq:physical_correlator}
\end{align}
The anti-holomorphic sector is identical up to $p_L^+ \to p_R^-$.

\begin{figure}%
    \centering
    {{\includegraphics[width=6cm]{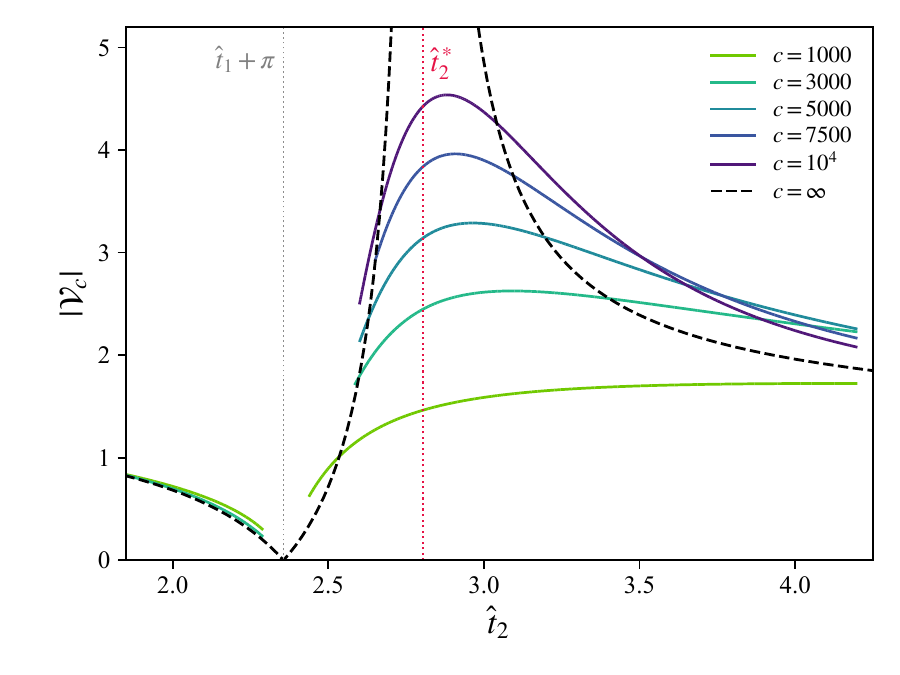} }}%
    \quad
    {{\includegraphics[width=6cm]{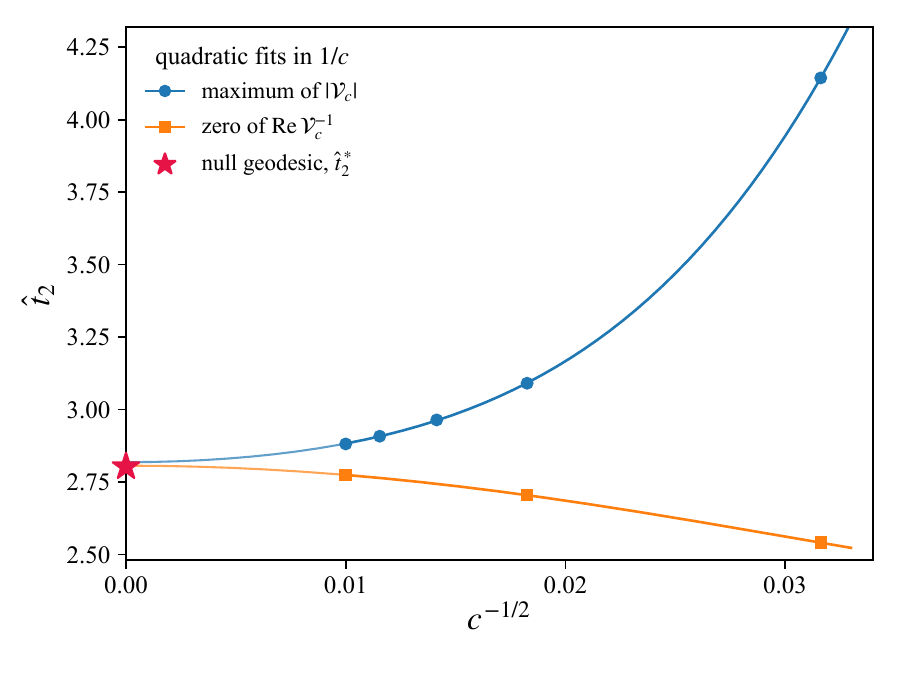} }}
    \caption{Example with $\epsilon=0.05$, $p = 1/2$, $h_L=1/2$, and $\hat t_1=-\pi/4$. Left: the normalized magnitude $|\mathcal V_c|$ for $c=1000,3000,5000,7500,10000$; the dashed curve is the matched non-zero $\epsilon$ two-defect classical result. Dotted lines indicate the boundary and bulk light cones at $\hat t_2=2.36$ and $2.80$.  Right: peak locations (circles) for all five charges and zeros of $\operatorname{Re}\mathcal V_c^{-1}$ (squares) for $c=1000,3000,10000$, plotted against $c^{-1/2}$. Both are fitted with a quadratic in $1/c$.}%
    \label{fig:exampleAhmed}%
\end{figure}

\paragraph{Resolution at finite $c$:} We evaluate the six-point function in the Virasoro comb channel $(\mathbf{1},L,\mathbf{1})$, where the two heavy pairs fuse to the identity which then fuses with the light primary $L$. The vacuum block is evaluated with Zamolodchikov's elliptic recursion \cite{Zamolodchikov}. We continue to the Lorentzian sheet by writing $\tau_1=\delta+i\lambda\hat t_1$ and $\tau_2=-\delta+i\lambda\hat t_2$ and going from $\lambda=0$ to $1$, then taking $\delta\to0^+$.  Denoting this normalized scalar channel contribution by $G_c$, define
\begin{equation}
    \mathcal V_c^2=\frac{G_c}{G_0},\qquad
    G_0=\left[4\cos^2\!\left(\frac{\hat t_2-\hat t_1}{2}\right)\right]^{-1},\qquad
    |\mathcal V_c|=\sqrt{\left|\frac{G_c}{G_0}\right|}.
    \label{eq:sixpoint_finitec_normalization}
\end{equation}
Here $G_0$ is the vacuum scalar probe two-point function for $h_L=\bar h_L=1/2$, and the branch of $\mathcal V_c$ is inherited from the same continuation.

We present the results in figure~\ref{fig:exampleAhmed}. The parameters used are
\begin{gather}
    \frac{h_H}{c}\approx 6.55\times10^{-4},\qquad
    \epsilon=0.05,\qquad h_L=\bar h_L=\frac12,\qquad
    \hat t_1=-\frac\pi4,
    \label{eq:sixpoint_finitec_parameters}
\end{gather}
with the probe endpoints at $\hat\varphi_1=0$ and $\hat\varphi_2=\pi$. These parameters are held fixed as we vary $c$. At these values, the strength of the shockwaves is $p = 12 \pi h_H/c\epsilon \approx 0.494$. As we indicate in the figure, we test $c$ ranging from $10^3$ to $10^4$, and confirm that the bulk cone singularity is reduced to a smooth peak. See the  caption of figure ~\ref{fig:exampleAhmed} for more details.

\paragraph{Resolution from quantum gravity:} We suggested in the introduction that metric quantum fluctuations might be responsible for resolving bulk cone singularities. We test this idea by showing that the superposition of geometries created by the two heavy operators reproduces the finite $c$ numerics to high accuracy.

\begin{figure}%
    \centering
    {{\includegraphics[width=6cm]{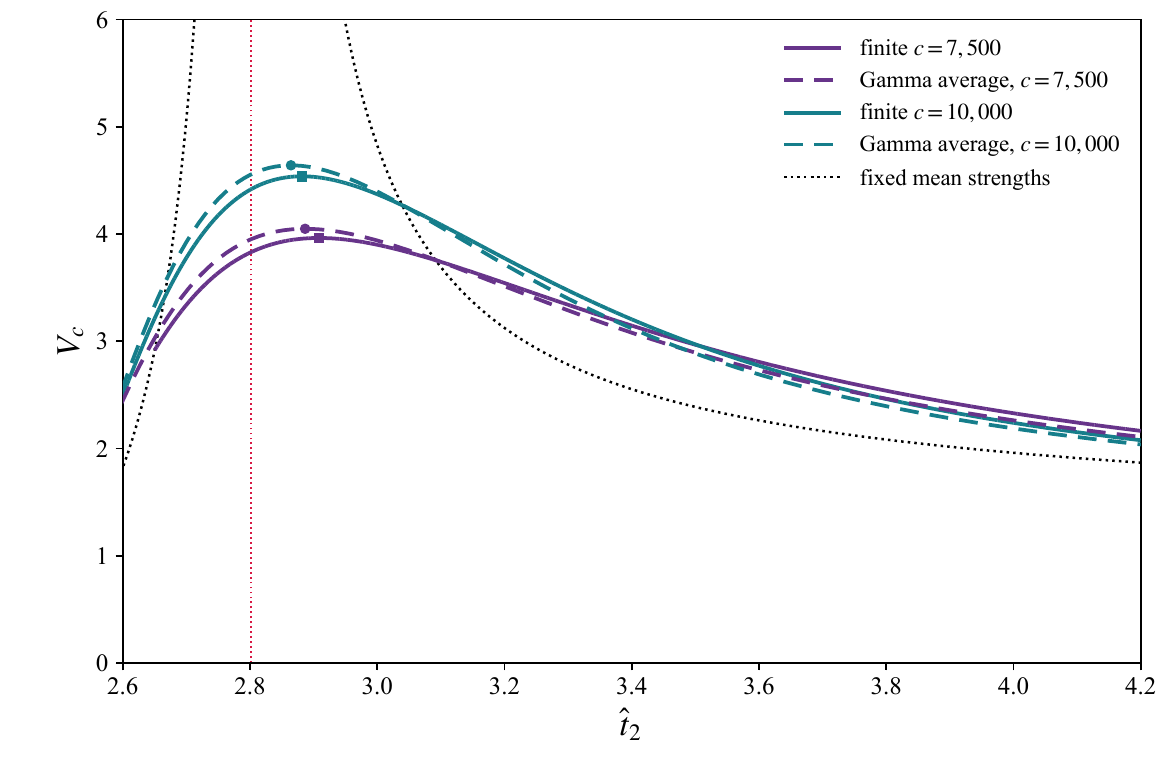} }}%
    \quad
    {{\includegraphics[width=6cm]{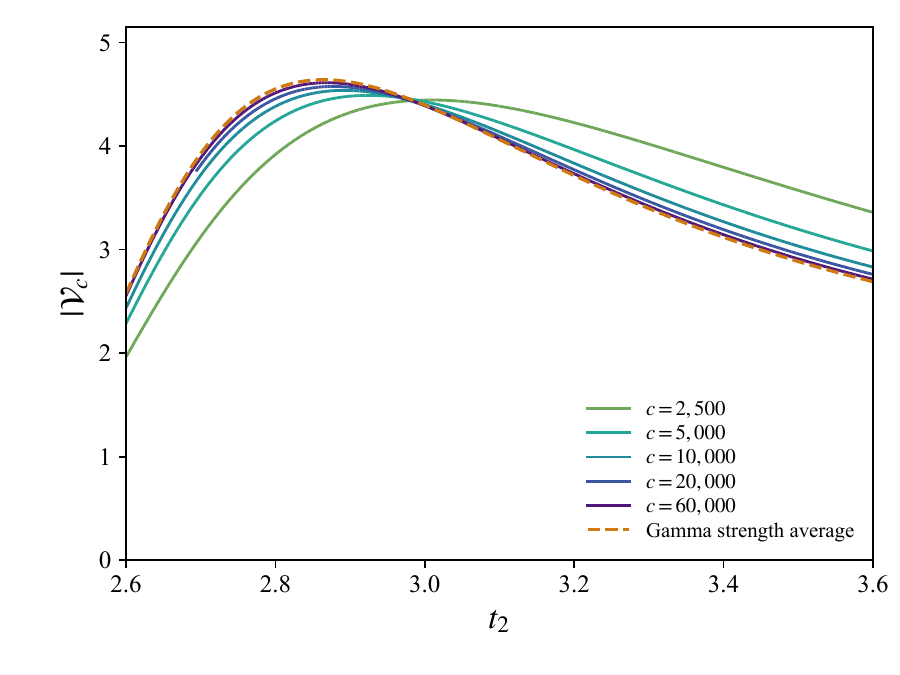} }}
    \caption{Left: There is strong agreement ($2\%$ peak difference) between the finite $c$ and the Gamma averaged infinite $c$ correlator where $h_H/c = 6.55 \times 10^{-4}$, $\epsilon = 0.05$, and $h_L = 1/2$. Right:  The finite $c$ computation appears to converge to the Gamma distribution average of the gravity computation at fixed $p = 1/2, \ h_H \approx 6.55$ in the $c \sim \epsilon^{-1} \to \infty$ limit.}%
    \label{fig:average}%
\end{figure}

The scattering defects state includes a pair of primary insertions a distance $-\epsilon$ in Euclidean time. This state prepares a superposition of geometries with different ADM energy and angular momentum. For a state with a single particle, the normalized two-point function gives the probability of finding the descendant state $L_{-1}^n|h_H \rangle$ to be
\begin{align}
    P_n = (1 - e^{-2 \epsilon})^{2h_H} {\Gamma(2h_H + n) \over \Gamma(2h_H)\Gamma(n + 1)} e^{-2 \epsilon n},
\end{align}
for one chiral family, and similarly for the other family. This distribution assigns the mean value $\langle h_H + n \rangle = h_H/\tanh \epsilon$. In the limit $c \to \infty, \epsilon \to 0$ keeping $h_H$ and $p = 12 \pi h_H/c \epsilon$ fixed, this discrete distribution approximates the Gamma distribution 
\begin{align}
    P(p) = \left( 2 h_H \over p_0 \right)^{2 h_H}{p^{2 h_H - 1}e^{-2ph_H/p_0} \over \Gamma(2 h_H)},
\end{align}
where $p_0 = 12\pi h_H/c\epsilon$ is the mean. We model summing over the geometries in the superposition by averaging over the shockwave strengths $p_R^\pm$ and $p_L^\pm$ independently using the Gamma distribution. The results are presented in figure \ref{fig:average}, showing qualitative agreement with the finite $c$ numerics.

We can see the smoothing of the singularity from a saddle point analysis of the averaging integral. For simplicity, consider the average of the holomorphic sector of the correlator \eqref{eq:physical_correlator}
\begin{align}
     {\left( 2 h_H / p_0 \right)^{2 h_H} \over \Gamma(2 h_H)}\int_{0}^\infty dp{p^{2 h_H - 1}e^{-2ph_H/p_0} \over  \left( \sin {u_2 - u_1 \over 2} - 2 p \cos {u_2 - \pi/2 \over 2}\cos {u_1 - \pi/2 \over 2} \right)^{2h_L}},
\end{align}
where $u_2$ is the arrival point and $u_1$ is the launching point. The saddle point equation at large $h_H$ reads
\begin{align}
    -{ h_H (p - p_0) \over p p_0} + {B(u_2,u_1)h_L \over A(u_2,u_1) - p B(u_2,u_1) } = 0,
\end{align}
where
\begin{align}
    A = \sin {u_2 - u_1 \over 2} , \quad B = 2 \cos {u_2 - \pi/2 \over 2}\cos {u_1 - \pi/2 \over 2}.
\end{align}
The classical bulk-cone singularity is located at $A  = p_0 B$ where $p_0$ is the mean of the distribution.  Near this point, however, the saddle point receives a large correction from the light operator correlator, despite $h_L \ll h_H$, and moves off into the complex plane
\begin{align}
    p_s = p_0 \left( 1 \pm {i \over \sqrt{2h_H}}\right).
\end{align}
Therefore, the saddle continues to move away as you approach it, and so the denominator in the integrand can never be tuned to vanish. The bulk cone singularity is then completely resolved. This mechanism is similar to the resolution of forbidden singularities in the microcanonical two-point function due to backreaction \cite{Bounce:FKLW,arXiv:1703.09727,arXiv:1712.03464,1809.08647}. One difference, however, is that there the saddle remains real, while here it becomes complex. It would be good to understand the connection between this complexification, the fuzziness introduced by quantum gravity and backreaction.

\section*{Acknowledgements}
AA would like to thank Shadi Ali Ahmad for useful conversations. We acknowledge the use of OpenAI’s Codex and Anthropic's Claude Code.  We used them for some editing, figures, and numerics. MD's work is supported by DOE grant DE-SC/0007870 and the Frankel-Goldfield Research Fund.

\bibliographystyle{JHEP}
\bibliography{references}

\providecommand{\href}[2]{#2}\begingroup\raggedright\begin{thebibliography}{10}

\bibitem{HubenyLiuRangamani:2006}
V.~E. Hubeny, H.~Liu, and M.~Rangamani, {\it {Bulk-cone singularities \&
  signatures of horizon formation in AdS/CFT}},  {\em JHEP} {\bf 01} (2007)
  009, [\href{http://arxiv.org/abs/hep-th/0610041}{{\tt hep-th/0610041}}].

\bibitem{DodelsonOoguri:2020}
M.~Dodelson and H.~Ooguri, {\it {Singularities of thermal correlators at strong
  coupling}},  {\em Phys. Rev. D} {\bf 103} (2021) 066018,
  [\href{http://arxiv.org/abs/2010.09734}{{\tt arXiv:2010.09734}}].

\bibitem{Sheta:StringyPhotonSphere}
A.~Sheta, ``{The stringy photon sphere}.'' to appear.

\bibitem{DodelsonEtAl:2023}
M.~Dodelson, C.~Iossa, R.~Karlsson, A.~Lupsasca, and A.~Zhiboedov, {\it {Black
  hole bulk-cone singularities}},  {\em JHEP} {\bf 07} (2024) 046,
  [\href{http://arxiv.org/abs/2310.15236}{{\tt arXiv:2310.15236}}].

\bibitem{Batra:2026otu}
G.~Batra, {\it {A non-prompt null ray cures its own singularity}},
  \href{http://arxiv.org/abs/2610.22478}{{\tt arXiv:2610.22478}}.

\bibitem{FidkowskiEtAl:2003}
L.~Fidkowski, V.~Hubeny, M.~Kleban, and S.~Shenker, {\it {The Black Hole
  Singularity in AdS/CFT}},  {\em JHEP} {\bf 02} (2004) 014,
  [\href{http://arxiv.org/abs/hep-th/0306170}{{\tt hep-th/0306170}}].

\bibitem{MaldacenaSimmonsDuffinZhiboedov:2015}
J.~Maldacena, D.~Simmons-Duffin, and A.~Zhiboedov, {\it {Looking for a bulk
  point}},  {\em JHEP} {\bf 01} (2017) 013,
  [\href{http://arxiv.org/abs/1509.03612}{{\tt arXiv:1509.03612}}].

\bibitem{FestucciaLiu:2008}
G.~Festuccia and H.~Liu, {\it {A Bohr--Sommerfeld quantization formula for
  quasinormal frequencies of AdS black holes}},  {\em Adv. Sci. Lett.} {\bf 2}
  (2009) 221--235, [\href{http://arxiv.org/abs/0811.1033}{{\tt
  arXiv:0811.1033}}].

\bibitem{Bounce:KOS}
P.~Kraus, H.~Ooguri, and S.~Shenker, {\it {Inside the horizon with AdS/CFT}},
  \href{http://arxiv.org/abs/hep-th/0212277}{{\tt hep-th/0212277}}.

\bibitem{Bounce:FL}
G.~Festuccia and H.~Liu, {\it {Excursions beyond the horizon: black hole
  singularities in Yang--Mills theories (I)}},
  \href{http://arxiv.org/abs/hep-th/0506202}{{\tt hep-th/0506202}}.

\bibitem{Bounce:FH}
A.~L. Fitzpatrick and K.-W. Huang, {\it {Universal Lowest-Twist in CFTs from
  Holography}},  \href{http://arxiv.org/abs/1903.05306}{{\tt
  arXiv:1903.05306}}.

\bibitem{Bounce:PSWambient}
E.~Parisini, K.~Skenderis, and B.~Withers, {\it {The Ambient Space Formalism}},
   \href{http://arxiv.org/abs/2312.03820}{{\tt arXiv:2312.03820}}.

\bibitem{Bounce:CLPV}
N.~{\v C}eplak, H.~Liu, A.~Parnachev, and S.~Valach, {\it {Black Hole
  Singularity from OPE}},  \href{http://arxiv.org/abs/2404.17286}{{\tt
  arXiv:2404.17286}}.

\bibitem{Bounce:Huang}
K.-W. Huang, {\it {Resummation of multistress tensors in higher dimensions}},
  \href{http://arxiv.org/abs/2406.07458}{{\tt arXiv:2406.07458}}.

\bibitem{Bounce:BGP1}
I.~Buri{\'c}, I.~Gusev, and A.~Parnachev, {\it {Thermal holographic correlators
  and KMS condition}},  \href{http://arxiv.org/abs/2505.10277}{{\tt
  arXiv:2505.10277}}.

\bibitem{Bounce:BGP2}
I.~Buri{\'c}, I.~Gusev, and A.~Parnachev, {\it {Holographic correlators from
  thermal bootstrap}},  \href{http://arxiv.org/abs/2508.08373}{{\tt
  arXiv:2508.08373}}.

\bibitem{Bounce:Valach}
S.~Valach, {\it {Thermal Correlators and Black Holes: From Infinity to
  Singularity}},  \href{http://arxiv.org/abs/2508.17139}{{\tt
  arXiv:2508.17139}}.

\bibitem{Bounce:CLPV2}
N.~{\v C}eplak, H.~Liu, A.~Parnachev, and S.~Valach, {\it {Fooling the censor:
  going beyond inner horizons with the OPE}},
  \href{http://arxiv.org/abs/2511.09638}{{\tt arXiv:2511.09638}}.

\bibitem{Bounce:CV}
N.~{\v C}eplak and S.~Valach, {\it {Where's Kasner? OPE Signature of Kasner
  Exponents}},  \href{http://arxiv.org/abs/2609.23776}{{\tt arXiv:2609.23776}}.

\bibitem{Bounce:AJCCM}
N.~Afkhami-Jeddi, S.~Caron-Huot, J.~Chakravarty, and A.~Maloney, {\it {Imprint
  of the black hole singularity on thermal two-point functions}},
  \href{http://arxiv.org/abs/2510.21673}{{\tt arXiv:2510.21673}}.

\bibitem{Bounce:Chak}
J.~Chakravarty, {\it {Imprint of black hole interior on thermal four-point
  correlators}},  \href{http://arxiv.org/abs/2512.10912}{{\tt
  arXiv:2512.10912}}.

\bibitem{Bounce:JR1}
H.~F. Jia and M.~Rangamani, {\it {Thermal spectral function asymptotics and
  black hole singularity in holography}},
  \href{http://arxiv.org/abs/2512.15114}{{\tt arXiv:2512.15114}}.

\bibitem{Bounce:JR2}
H.~F. Jia and M.~Rangamani, {\it {Exact holographic thermal spectral functions:
  OPE, non-perturbative corrections, and black hole singularity}},
  \href{http://arxiv.org/abs/2604.10803}{{\tt arXiv:2604.10803}}.

\bibitem{Bounce:AW1}
P.~Arnaudo and B.~Withers, {\it {Analytic structure of holographic thermal
  correlators from Fourier series}},
  \href{http://arxiv.org/abs/2603.13469}{{\tt arXiv:2603.13469}}.

\bibitem{Bounce:GLS}
S.~Giombi, Y.-Z. Li, and J.~Shan, {\it {Bouncing singularities and thermal
  correlators on line defects}},  \href{http://arxiv.org/abs/2603.11012}{{\tt
  arXiv:2603.11012}}.

\bibitem{Bounce:AH}
I.~Amado and C.~Hoyos-Badajoz, {\it {AdS black holes as reflecting cavities}},
  \href{http://arxiv.org/abs/0807.2337}{{\tt arXiv:0807.2337}}.

\bibitem{Bounce:HLQZ}
G.~T. Horowitz, H.~Leung, L.~Queimada, and Y.~Zhao, {\it {Boundary signature of
  singularity in the presence of a shock wave}},
  \href{http://arxiv.org/abs/2310.03076}{{\tt arXiv:2310.03076}}.

\bibitem{Bounce:GVV}
S.~Grozdanov, S.~Valach, and M.~Vrbica, {\it {Bouncing geodesics, black hole
  singularities, and singularities of thermal correlators}},
  \href{http://arxiv.org/abs/2603.15598}{{\tt arXiv:2603.15598}}.

\bibitem{Bounce:JK}
Y.~Jia and M.~Kulaxizi, {\it {Bulk Phase Shift and Singularity}},
  \href{http://arxiv.org/abs/2602.06558}{{\tt arXiv:2602.06558}}.

\bibitem{Bounce:AEJKP}
I.~J. Araya, C.~Esper, Y.~Jia, M.~Kulaxizi, and A.~Parnachev, {\it {Bulkcone
  Singularities and Complex Geodesics}},
  \href{http://arxiv.org/abs/2602.12893}{{\tt arXiv:2602.12893}}.

\bibitem{Bounce:AAL}
S.~Ali~Ahmad, A.~Almheiri, and S.~Lin, {\it {Continuing past the inner horizon
  using WKB}},  \href{http://arxiv.org/abs/2601.02354}{{\tt arXiv:2601.02354}}.

\bibitem{Bounce:GMV}
S.~Grozdanov, V.~Movrin, and S.~Valach, {\it {Bouncing Geodesics,
  Singularities, and the Cavity Thermal Product Formula in Asymptotically Flat
  and de Sitter Black Holes}},  \href{http://arxiv.org/abs/2606.11297}{{\tt
  arXiv:2606.11297}}.

\bibitem{Bounce:DIKZ}
M.~Dodelson, C.~Iossa, R.~Karlsson, and A.~Zhiboedov, {\it {A thermal product
  formula}},  \href{http://arxiv.org/abs/2304.12339}{{\tt arXiv:2304.12339}}.

\bibitem{Bounce:AW2}
P.~Arnaudo and B.~Withers, {\it {Bouncing singularities in Schwarzschild: a
  geometric origin of the QNM convergence region}},
  \href{http://arxiv.org/abs/2605.16489}{{\tt arXiv:2605.16489}}.

\bibitem{Bounce:HZ}
S.~A. Hartnoll and A.~Zhiboedov, {\it {Can one hear the shape of a black hole
  singularity?}},  \href{http://arxiv.org/abs/2609.07514}{{\tt
  arXiv:2609.07514}}.

\bibitem{Bounce:FrenkelKasner}
A.~Frenkel, S.~A. Hartnoll, J.~Kruthoff, and Z.~D. Shi, {\it {Holographic flows
  from CFT to the Kasner universe}},  {\em JHEP} {\bf 08} (2020) 003,
  [\href{http://arxiv.org/abs/2004.01192}{{\tt arXiv:2004.01192}}].

\bibitem{Bounce:CGHSSquantum}
C.~Coviello, A.~Gupta, R.~A. Hennigar, K.~Shi, and A.~Svesko, {\it {Looking
  inside a quantum black hole}},  \href{http://arxiv.org/abs/2609.20788}{{\tt
  arXiv:2609.20788}}.

\bibitem{Bounce:CGHSSinteriors}
C.~Coviello, A.~Gupta, R.~A. Hennigar, K.~Shi, and A.~Svesko, {\it {Quasinormal
  modes and quantum black hole interiors}},
  \href{http://arxiv.org/abs/2609.26724}{{\tt arXiv:2609.26724}}.

\bibitem{Bounce:DIK}
M.~Dodelson, C.~Iossa, and R.~Karlsson, {\it {Bouncing off a stringy
  singularity}},  \href{http://arxiv.org/abs/2511.09616}{{\tt
  arXiv:2511.09616}}.

\bibitem{Bounce:BCGHPR}
I.~Buri{\'c}, C.-M. Chang, I.~Gusev, E.~Helfenberger, A.~Parnachev, and
  M.~Rangamani, {\it {Thermal two-point functions in SYK and complex-time
  singularities}},  \href{http://arxiv.org/abs/2607.05258}{{\tt
  arXiv:2607.05258}}.

\bibitem{Bounce:FKLW}
A.~L. Fitzpatrick, J.~Kaplan, D.~Li, and J.~Wang, {\it {On Information Loss in
  AdS$_3$/CFT$_2$}},  \href{http://arxiv.org/abs/1603.08925}{{\tt
  arXiv:1603.08925}}.

\bibitem{Bounce:GM}
M.~Grinberg and J.~Maldacena, {\it {Proper time to the black hole singularity
  from thermal one-point functions}},
  \href{http://arxiv.org/abs/2011.01004}{{\tt arXiv:2011.01004}}.

\bibitem{Bounce:RGRsingularity}
D.~Rodriguez-Gomez and J.~G. Russo, {\it {Correlation functions in finite
  temperature CFT and black hole singularities}},
  \href{http://arxiv.org/abs/2102.11891}{{\tt arXiv:2102.11891}}.

\bibitem{Bounce:KRthermal}
H.~Krishna and D.~Rodriguez-Gomez, {\it {Holographic thermal correlators
  revisited}},  \href{http://arxiv.org/abs/2108.00277}{{\tt arXiv:2108.00277}}.

\bibitem{Bounce:BMonepoint}
D.~Berenstein and R.~Mancilla, {\it {Aspects of thermal one-point functions and
  response functions in AdS Black holes}},
  \href{http://arxiv.org/abs/2211.05144}{{\tt arXiv:2211.05144}}.

\bibitem{Bounce:DK}
J.~R. David and S.~Kumar, {\it {Thermal one point functions, large $d$ and
  interior geometry of black holes}},
  \href{http://arxiv.org/abs/2212.07758}{{\tt arXiv:2212.07758}}.

\bibitem{Bounce:Singhi}
K.~Singhi, {\it {Proper time to the singularity and thermal correlators}},
  \href{http://arxiv.org/abs/2406.08553}{{\tt arXiv:2406.08553}}.

\bibitem{Bounce:GKPdS}
A.~Goldar, N.~Kajuri, and R.~Pal, {\it {Holographic One-Point Functions and
  Geodesics in SdS$_3$}},  \href{http://arxiv.org/abs/2603.21425}{{\tt
  arXiv:2603.21425}}.

\bibitem{Bounce:BBMMPholography}
J.~Barrat, D.~N. Bozkurt, E.~Marchetto, A.~Miscioscia, and E.~Pomoni, {\it
  {Analytic thermal bootstrap meets holography}},
  \href{http://arxiv.org/abs/2510.20894}{{\tt arXiv:2510.20894}}.

\bibitem{Komargodski:2026Bouncing}
Z.~Komargodski, {\it {Fate of Bouncing Singularities at Finite $G_N$}},
  \href{http://arxiv.org/abs/2609.31309}{{\tt arXiv:2609.31309}}.

\bibitem{BatraLevineShenker:2026}
G.~Batra, A.~Levine, and S.~H. Shenker, {\it {A backreacted bouncing geodesic
  in 2D}},  \href{http://arxiv.org/abs/2609.32071}{{\tt arXiv:2609.32071}}.

\bibitem{Pappadopulo:2012jk}
D.~Pappadopulo, S.~Rychkov, J.~Espin, and R.~Rattazzi, {\it {OPE Convergence in
  Conformal Field Theory}},  {\em Phys. Rev. D} {\bf 86} (2012) 105043,
  [\href{http://arxiv.org/abs/1208.6449}{{\tt arXiv:1208.6449}}].

\bibitem{Hogervorst:2013sma}
M.~Hogervorst and S.~Rychkov, {\it {Radial Coordinates for Conformal Blocks}},
  {\em Phys. Rev. D} {\bf 87} (2013) 106004,
  [\href{http://arxiv.org/abs/1303.1111}{{\tt arXiv:1303.1111}}].

\bibitem{Kravchuk:2021kwe}
P.~Kravchuk, J.~Qiao, and S.~Rychkov, {\it {Distributions in CFT. Part II.
  Minkowski space}},  {\em JHEP} {\bf 08} (2021) 094,
  [\href{http://arxiv.org/abs/2104.02090}{{\tt arXiv:2104.02090}}].

\bibitem{HartmanJainKundu:2015}
T.~Hartman, S.~Jain, and S.~Kundu, {\it {Causality Constraints in Conformal
  Field Theory}},  {\em JHEP} {\bf 05} (2016) 099,
  [\href{http://arxiv.org/abs/1509.00014}{{\tt arXiv:1509.00014}}].

\bibitem{Leung:2024}
H.~Leung, {\it {New bulk cone singularities in Vaidya-like spacetimes from
  large $c$ conformal blocks}},  \href{http://arxiv.org/abs/2411.19924}{{\tt
  arXiv:2411.19924}}.

\bibitem{Anous:2016}
T.~Anous, T.~Hartman, A.~Rovai, and J.~Sonner, {\it {Black Hole Collapse in the
  $1/c$ Expansion}},  {\em JHEP} {\bf 07} (2016) 123,
  [\href{http://arxiv.org/abs/1603.04856}{{\tt arXiv:1603.04856}}].

\bibitem{Anous:2017}
T.~Anous, T.~Hartman, A.~Rovai, and J.~Sonner, {\it {From Conformal Blocks to
  Path Integrals in the Vaidya Geometry}},  {\em JHEP} {\bf 09} (2017) 009,
  [\href{http://arxiv.org/abs/1706.02668}{{\tt arXiv:1706.02668}}].

\bibitem{BerensteinGrabovskyLi:2023}
D.~Berenstein, D.~Grabovsky, and Z.~Li, {\it {Aspects of Holography in Conical
  AdS$_3$}},  {\em JHEP} {\bf 04} (2023) 029,
  [\href{http://arxiv.org/abs/2205.02256}{{\tt arXiv:2205.02256}}].

\bibitem{Grabovsky:2024}
D.~Grabovsky, {\it {Heavy States in 3d Gravity and 2d CFT}},  {\em JHEP} {\bf
  07} (2024) 287, [\href{http://arxiv.org/abs/2403.13757}{{\tt
  arXiv:2403.13757}}].

\bibitem{banados}
M.~Ba{\~n}ados, {\it {Three-dimensional quantum geometry and black holes}},
  {\em AIP Conf. Proc.} {\bf 484} (1999) 147--169,
  [\href{http://arxiv.org/abs/hep-th/9901148}{{\tt hep-th/9901148}}].

\bibitem{Matschull:1998rv}
H.-J. Matschull, {\it {Black hole creation in 2+1 dimensions}},  {\em Class.
  Quant. Grav.} {\bf 16} (1999) 1069,
  [\href{http://arxiv.org/abs/gr-qc/9809087}{{\tt gr-qc/9809087}}].

\bibitem{Lindgren:2016wtw}
E.~J. Lindgren, {\it {Collisions of massive particles, timelike thin shells and
  formation of black holes in three dimensions}},  {\em JHEP} {\bf 12} (2016)
  048, [\href{http://arxiv.org/abs/1611.02973}{{\tt arXiv:1611.02973}}].

\bibitem{GaoWald:2000}
S.~Gao and R.~M. Wald, {\it {Theorems on gravitational time delay and related
  issues}},  {\em Class. Quant. Grav.} {\bf 17} (2000) 4999--5008,
  [\href{http://arxiv.org/abs/gr-qc/0007021}{{\tt gr-qc/0007021}}].

\bibitem{FeffermanGraham:2012}
C.~Fefferman and C.~R. Graham, {\em {The Ambient Metric}}, vol.~178 of {\em
  Annals of Mathematics Studies}.
\newblock Princeton University Press, 2012.

\bibitem{FKW}
A.~L. Fitzpatrick, J.~Kaplan, and M.~T. Walters, {\it {Virasoro Conformal
  Blocks and Thermality from Classical Background Fields}},
  \href{http://arxiv.org/abs/1501.05315}{{\tt arXiv:1501.05315}}.

\bibitem{Zamolodchikov}
{\relax{}Al}.~B. Zamolodchikov, {\it {Conformal symmetry in two-dimensional
  space: recursion representation of conformal block}},  {\em Theor. Math.
  Phys.} {\bf 73} (1987) 1088.

\bibitem{arXiv:1703.09727}
H.~Chen, C.~Hussong, J.~Kaplan, and D.~Li, {\it {A Numerical Approach to
  Virasoro Blocks and the Information Paradox}},  {\em JHEP} {\bf 09} (2017)
  102, [\href{http://arxiv.org/abs/1703.09727}{{\tt arXiv:1703.09727}}].

\bibitem{arXiv:1712.03464}
T.~Faulkner and H.~Wang, {\it {Probing beyond ETH at large $c$}},  {\em JHEP}
  {\bf 06} (2018) 123, [\href{http://arxiv.org/abs/1712.03464}{{\tt
  arXiv:1712.03464}}].

\bibitem{1809.08647}
Z.~Yang, {\it {The Quantum Gravity Dynamics of Near Extremal Black Holes}},
  {\em JHEP} {\bf 05} (2019) 205, [\href{http://arxiv.org/abs/1809.08647}{{\tt
  arXiv:1809.08647}}].

\end{thebibliography}\endgroup

\end{document}